\documentclass{article}

\usepackage{graphicx}
\usepackage{graphics}
\usepackage{latexsym}
\usepackage{amsfonts}
\usepackage{amssymb}

\usepackage{amsmath}       
\usepackage{url}
\usepackage{subfigure}

\begin{document}

\title{Simple networks on complex cellular automata:\\ From de Bruijn diagrams to jump-graphs}

\author{Genaro J. Mart{\'i}nez$^{1,2}$, Andrew Adamatzky$^{2}$, \\ Bo Chen$^{3}$, Fangyue Chen$^{3}$, Juan C.S.T. Mora$^{4}$}

\date{25 November 2017\footnote{Chapter published in the book {\it Swarm Dynamics as a Complex Network}, Springer, (Ivan Zelinka and Guanrong Chen Eds.), chapter 12, pages 241-264, 2017.}}

\maketitle

\begin{centering}
$^1$ Escuela Superior de C\'omputo, Instituto Polit\'ecnico Nacional, M\'exico  \\
$^2$ International Centre of Unconventional Computing, University of the West of England, Bristol, United Kingdom \\
$^3$ School of Science, Hangzhou Dianzi University, Hangzhou, China \\
$^4$ \'Area Acad\'emica de Ingenier{\'i}a, Universidad Aut\'onoma del Estado de Hidalgo, Pachuca, Hidalgo, M\'exico  \\
\end{centering}

\begin{abstract}
\noindent
We overview networks which characterise dynamics in cellular automata. These networks are derived from  one-dimensional cellular automaton rules and global states of the automaton evolution: de Bruijn diagrams, subsystem diagrams, basins of attraction, and jump-graphs. These networks are used to understand properties of spatially-extended dynamical systems: emergence of non-trivial patterns, self-organisation, reversibility and chaos. Particular attention is paid to networks determined by travelling self-localisations, or gliders.
\end{abstract}

\section{Introduction}
\label{introduction}

Cellular automata (CA) are arrays of finite state machines,  or cells. Each cell takes a finite number of states. All cells update their states by the same rule simultaneously. A cell updates its state depending on states of its immediate neighbours. The CA originated from Stanislaw Ulam problem on a parallel transformation of matrices, and further developed in a context of self-reproduction by John von Neumann~\cite{von}. CA are apparently simple systems yet exhibiting sophisticated patterns of non-trivial behaviour. They are now ubiquitous tools in studying complex systems and non-linear dynamics in literally all fields of science and engineering, see e.g. a visual guide of representative power of CA in~\cite{artca}. Rules space and global dynamics of CA are characterised by networks: the de Bruijn diagrams \cite{mcbook, cavoorhees}, subsystem diagrams \cite{Guan, BoChen}, basins of attraction \cite{kn:WL92}, and jump-graphs \cite{ddlabbook}. We analyse predictive power of the networks using two CA rules which exhibit complex space-time dynamics: gliders, particles, waves, or localisations. The CA rules studied are the Rule 54\footnote{Repository Rule 54 \url{http://uncomp.uwe.ac.uk/genaro/Rule54.html}} and the Rule 110\footnote{Repository Rule 110 \url{http://uncomp.uwe.ac.uk/genaro/Rule110.html}}.

\section{Elementary cellular automata}
\label{eca}

A cell in elementary CA (ECA) takes two states from  $\Sigma=\{0,1\}$. A cell $x_i$, $1 \leq i \leq n$, updates its state by local function $\varphi$ depending on its own state and states of its two immediate neighbours:

$$
x_i^{t+1} = \varphi(x_{i-1}^t, x_i^t, x_{i+1}^t) .
$$

There are $2^{2^3}=256$ cell state transition rules. In 1983, Stephen Wolfram established a classification ECA rules \cite{cac}, based on space-time development of automata governed by the rules. The Wolfram classes are

\begin{itemize}
\item[] {\bf Class I}: evolution to {\it uniform} behaviour;
\item[] {\bf Class II}: evolution to {\it periodic} behaviour;
\item[] {\bf Class III}: evolution to {\it chaotic} behaviour;
\item[] {\bf Class IV}: evolution to {\it complex} behaviour.
\end{itemize}

Wolfram's classification is not the only one, there are 17 classifications of CA rule space~\cite{kn:Mar13}, however it is fully adequate and it became a classic `item' of CA theory.

The class III includes rules 54 and 110 (decimal representation of a binary rule-string). Rules in class III are called `complex' because the automata governed by these rules have longer transient periods and more sensitive to initial conditions than automata governed by rules form other classes. And, most importantly, they exhibit a wide range of travelling and stationary localisations, interactions that are the reason of `complexity' of the rules' behaviour.\footnote{Complex Cellular Automata Repository \url{http://uncomp.uwe.ac.uk/genaro/Complex_CA_repository.html}.} 

\begin{figure}
\includegraphics[scale=.385]{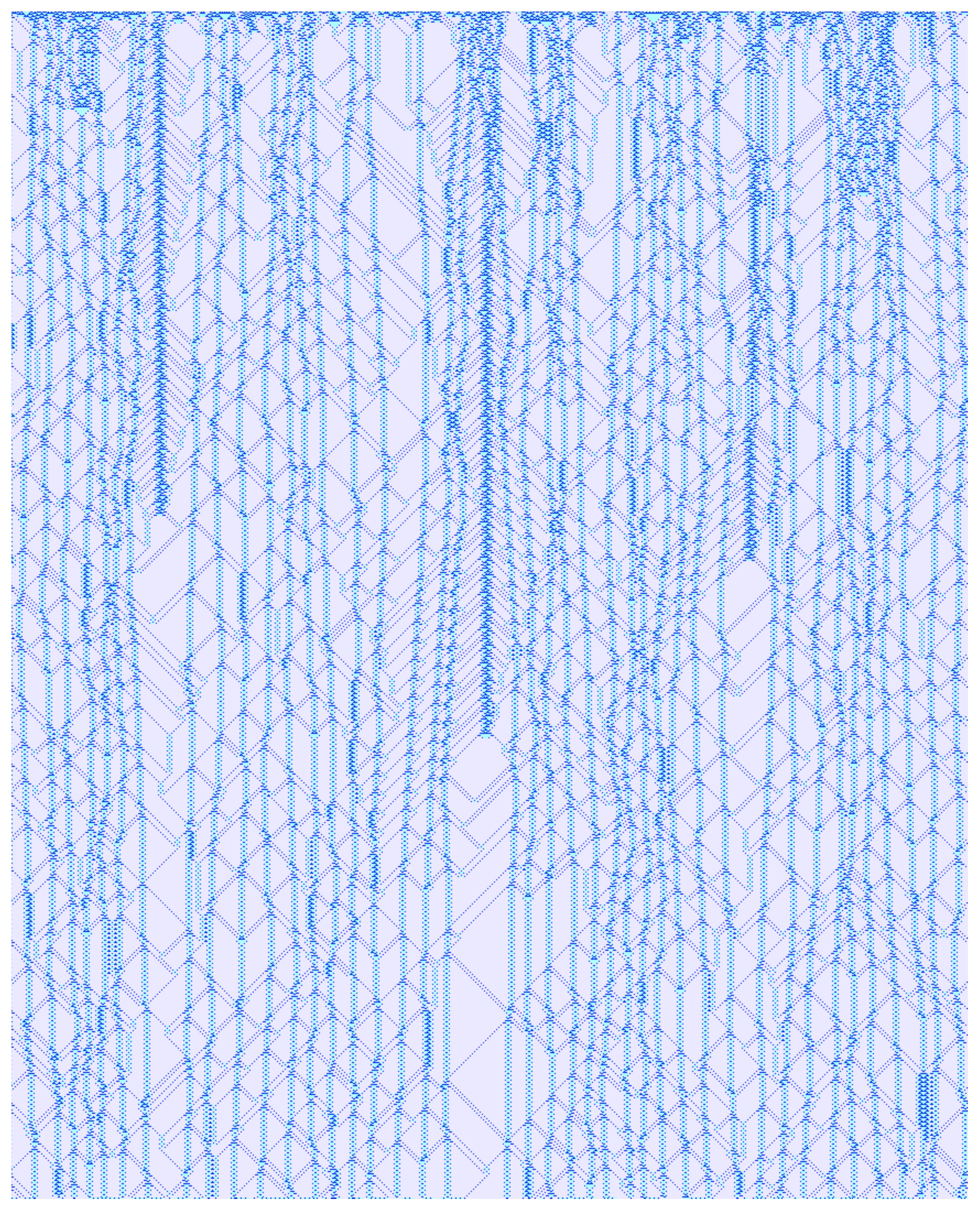}
\caption{Complex behaviour derived from ECA Rule 54. Non-trivial patterns as mobile localisations emerge and collide in the system. This snapshot evolves from a random initial configuration of 869 cells to 1,078 generations at 50\% of density. A filter is selected in the evolution for a better view of localisations and interactions.}
\label{ECAR54_869x1078}
\end{figure}

\begin{figure}
\includegraphics[scale=.385]{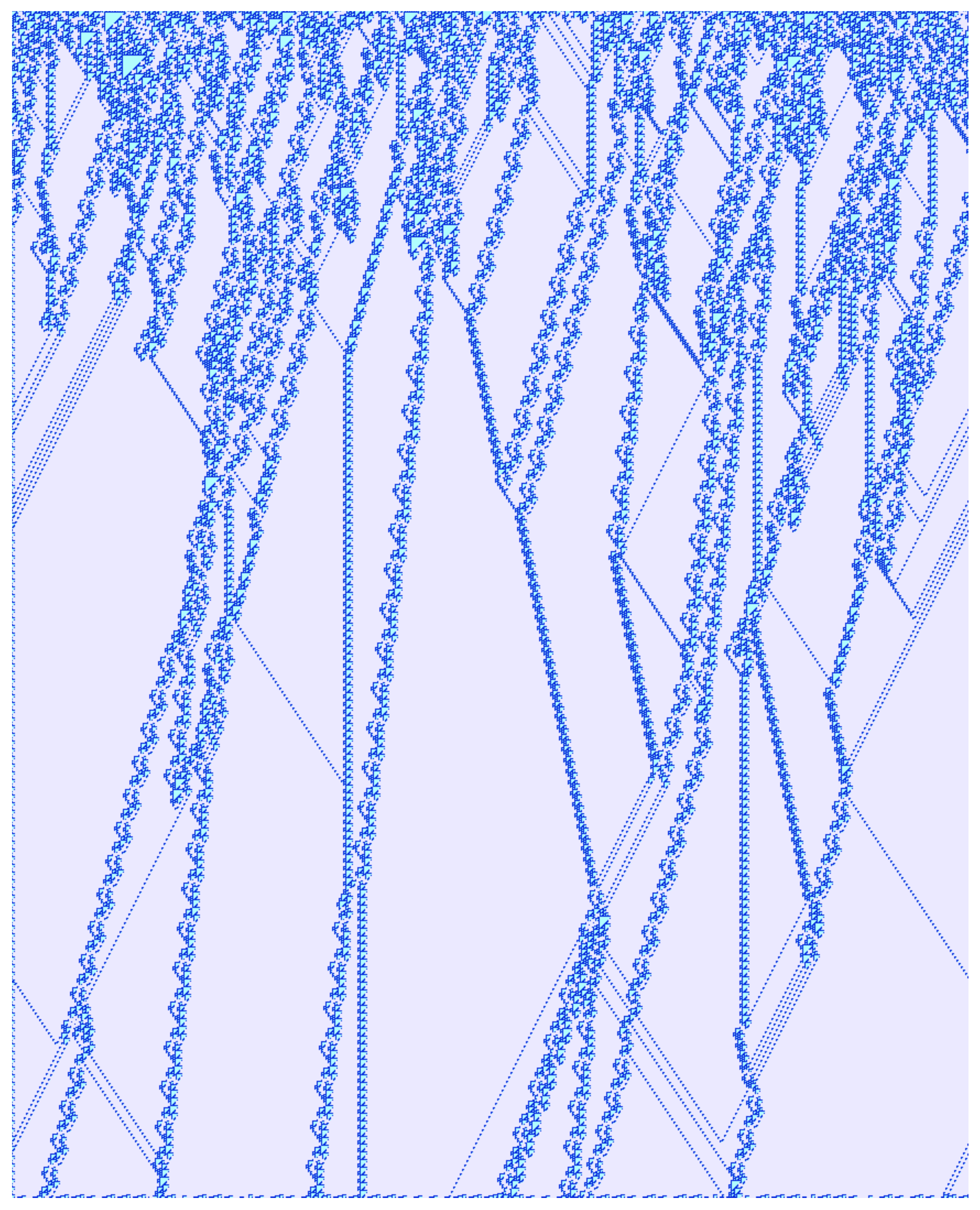}
\caption{Complex behaviour derived from ECA Rule 110. Non-trivial patterns as mobile localisations emerge and collide in the system. This snapshot evolves from a random initial configuration of 869 cells to 1,078 generations at 50\% of density. A filter is selected in the evolution for a better view of localisations and interactions.}
\label{ECAR110_869x1078}
\end{figure}

We study two ECA complex rules: Rules 54 and 110.  

The Rule 54 (`54' is a decimal representation of a binary string 00110110) automaton typically exhibits a rich dynamics of mobile localisations and outcomes of their collisions. Some of the mobile localisation collisions were used to demonstrate logical universality of the rule in~\cite{rule54}. The Rule 54 is represented by a function:

\begin{equation}
\varphi_{R54} = \left\{
	\begin{array}{lcl}
		1 & \mbox{if} & 101, 100, 010, 001 \\
		0 & \mbox{if} & 111, 110, 011, 000
	\end{array} \right. .
\label{functionR54}
\end{equation}

Figure~\ref{ECAR54_869x1078} shows a typical evolution of ECA Rule 54 from a random initial condition.

Rule 110 (`110' is a decimal representation of a binary string 01101110) is proved to be  computationally universal because it can simulate a cyclic tag system, which in turns simulates a universal Turing machine~\cite{ecauniversality}. The cell-state transition function of the Rule 110 is

\begin{equation}
\varphi_{R110} = \left\{
	\begin{array}{lcl}
		1 & \mbox{if} & 110, 101, 011, 010, 001 \\
		0 & \mbox{if} & 111, 100, 000
	\end{array} \right. .
\label{functionR110}
\end{equation}

Figure~\ref{ECAR110_869x1078} shows a typical evolution of ECA Rule 110 from a random initial condition.

\begin{figure}[th]
\begin{center}
\subfigure[]{\scalebox{0.3}{\includegraphics{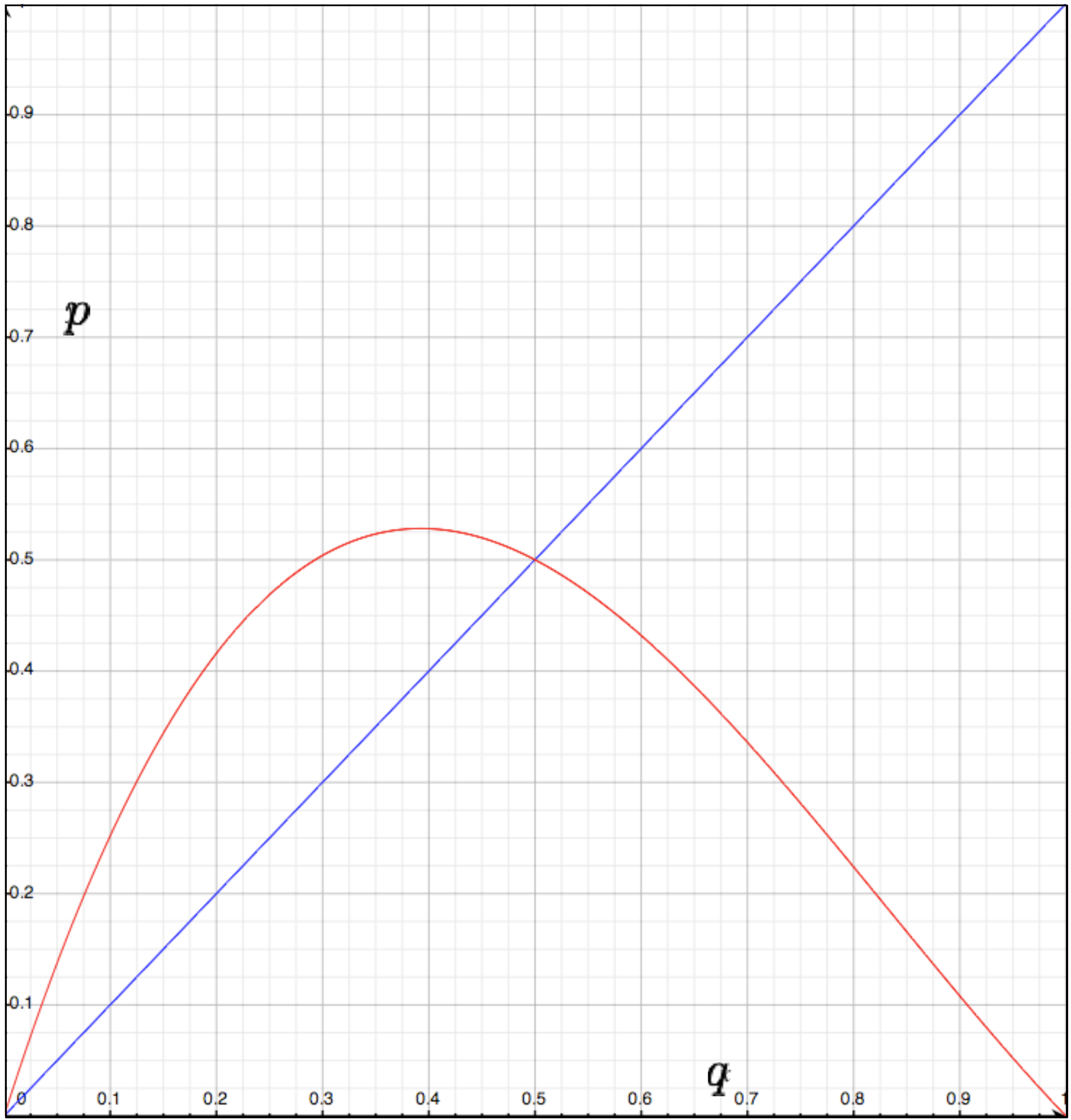}}} 
\subfigure[]{\scalebox{0.3}{\includegraphics{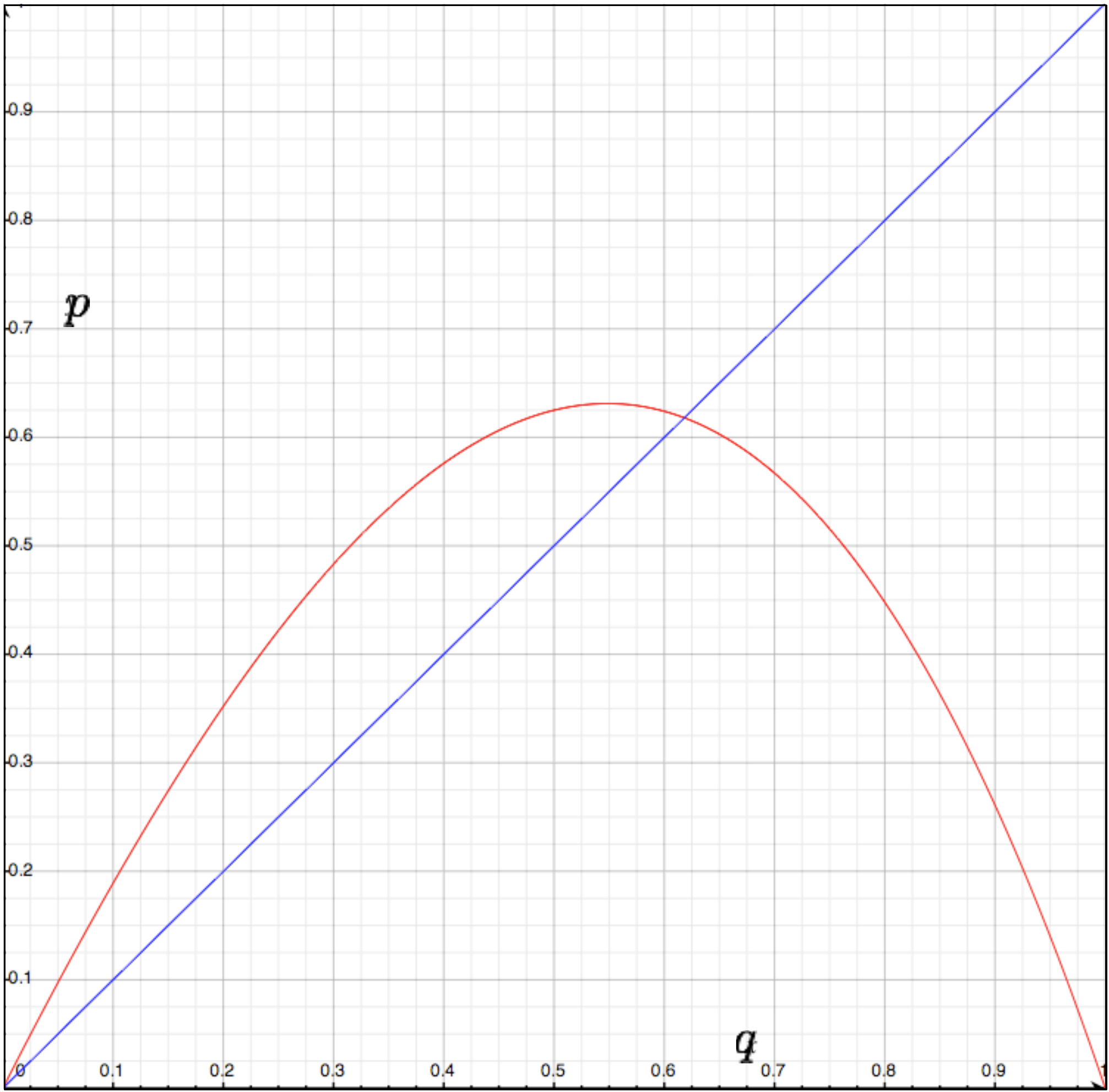}}}
\end{center}
\caption{Mean field curves for (a) Rule 54 and (b) Rule 110.}
\label{mfp}
\end{figure}

Complexity of CA rules can be estimated using polynomial approximation borrowed from the mean field theory~\cite{compcaclass}. Mean field theory techniques are efficient in discovering statistical properties of CA without analysing the whole evolution space of individual rules. These techniques have been introduced in CA field by Howard Gutowitz in \cite{kn:GVK87}. The approach assumes that cell-states do not correlate with each other in the local function $\varphi$. Thus we can study probabilities of states in a neighbourhood in terms of the probability of a single cell-state, and the probability of a neighbourhood would be the product of the probabilities of each cell in it. For one-dimensional CA with $k$-cell neighbourhood or radius $r$ and $k$ cell-states the probability is calculated as follows:

\begin{equation}
p_{t+1}=\sum_{j=0}^{q^{2r+1}-1}\varphi_{j}(X)p_{t}^{v}(1-p_{t})^{n-v}
\label{MFp1D}
\end{equation}

\noindent where  $j$ of a neighbourhood state, 
$q$ is a number of cell-states, 
$X$ is a neighbourhood $x_{i-r}, \ldots, x_{i}, \ldots, x_{i+r}$, 
$k$ is the number of cells in every neighbourhood, 
$v$ indicates how often state `1' occurs in $X$, 
$n-v$ shows how often state `0' occurs in the neighbourhood $X$, 
$p_{t}$ is the probability of a cell being in state `1', and $q_{t}$ is the probability of a cell being in state `0'; i.e., $q=1-p$.
In our case of binary cell-states and three-cell neighbourhood the probability is 
$$
p_{t+1} = \sum_{j=0}^{7} \varphi_{j}(X)p_{t}^{v}(1-p_{t})^{n-v}
$$

In \cite{mc90}, Harold McIntosh proposed a classification based on curves derived with the mean field approximation. A complex rule has a curve tangential to the identity, and an unstable fixed point that defines regions with unpredictable behaviour: 

\begin{center}
{\it Class IV: mean field curve horizontal plus diagonal tangency (no crossing the identity, possibly complex dynamics)}.
\end{center}

Figure~\ref{mfp}a shows a mean field curve for Rule 54 with a polynomial defined as:

\begin{equation}
p_{t+1} = 3p_tq^2_t+p^2_tq_t.
\end{equation}

The origin value is a stable fixed point which guarantees the stable configuration in state zero. The maximum point $p=0.5281$ is very close to the fixed stable point in $p=0.5$. Complex dynamics in Rule 54 emerges on a periodic background with the same number of states zero and one, thus the stable fixed point well characterises the local function (see Eq.~\ref{functionR54}). Also, this fixed point shows that a Rule 54 automaton starting from low or high densities of state 1 cells, more likely will finish its evolution with the same ratio of states (Fig.~\ref{ECAR54_869x1078}).

Figure~\ref{mfp}b shows a mean field curve for Rule 110 with polynomial defined as:

\begin{equation}
p_{t+1} = 2p_tq^{2}_t+3p^{2}_tq_t.
\end{equation}

The maximum point $p=0.6311$ is close to the fixed stable point in $p=0.62$. In Rule 110 we cannot find unstable fixed points. Complex dynamics in Rule 110 emerge quickly in few steps with a large number of mobile localisations and several chaotic regions. The fixed point characterises a periodic background that is often reached by Rule 110 automata 300 to 500 steps of development. Although, the number of states 1 is close to 0.57, the fixed point is reached because commonly an evolution in Rule 110 frequently is accompanied by mobile self-localisations traveling to the left (Fig.~\ref{ECAR110_869x1078}). For full details about mobile localisations in Rule 110 see \cite{glidersIJUC, reglangr110}.

\section{De Bruijn diagrams}

\begin{figure}
\begin{center}
\subfigure[]{\scalebox{0.35}{\includegraphics{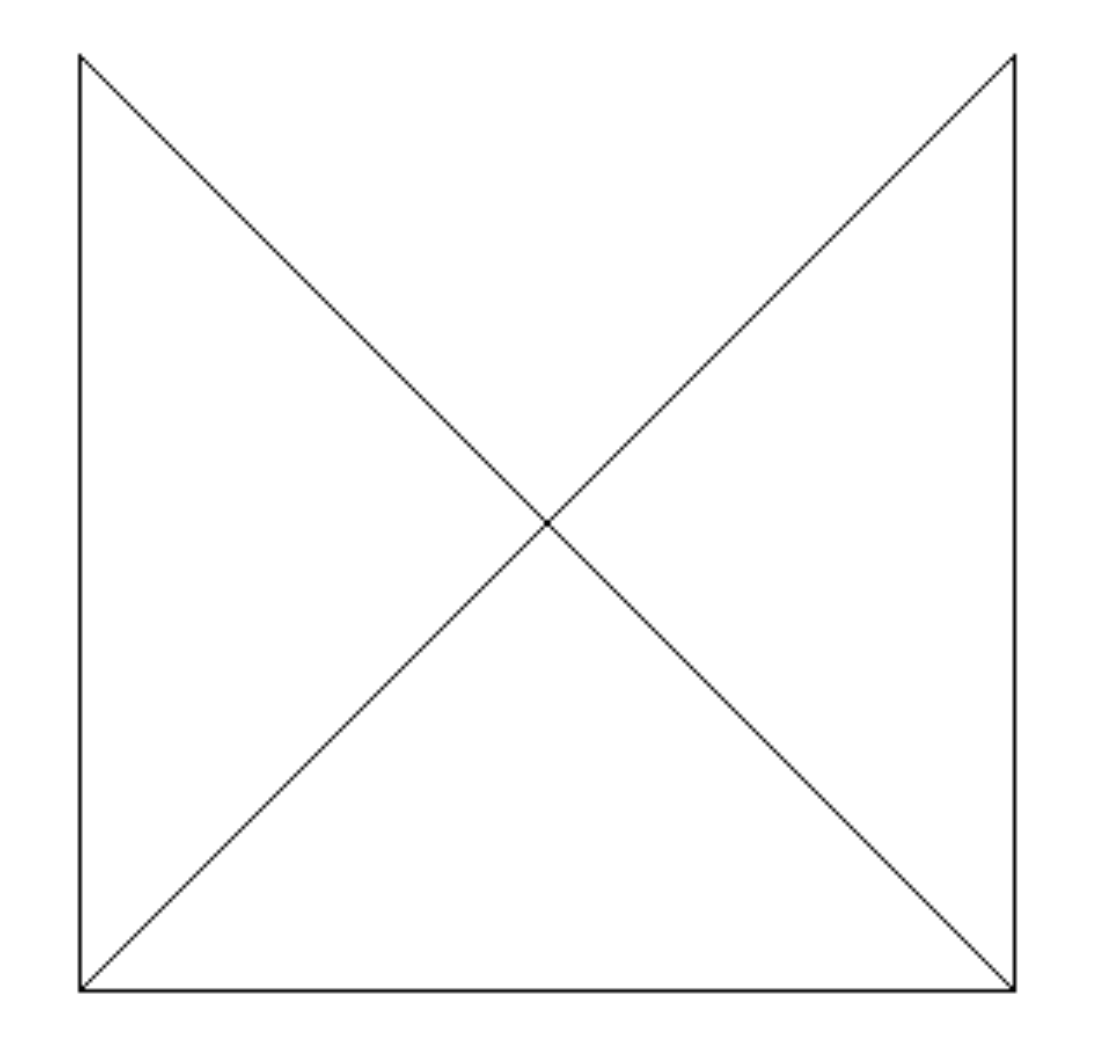}}} \hspace{1.4cm}
\subfigure[]{\scalebox{0.35}{\includegraphics{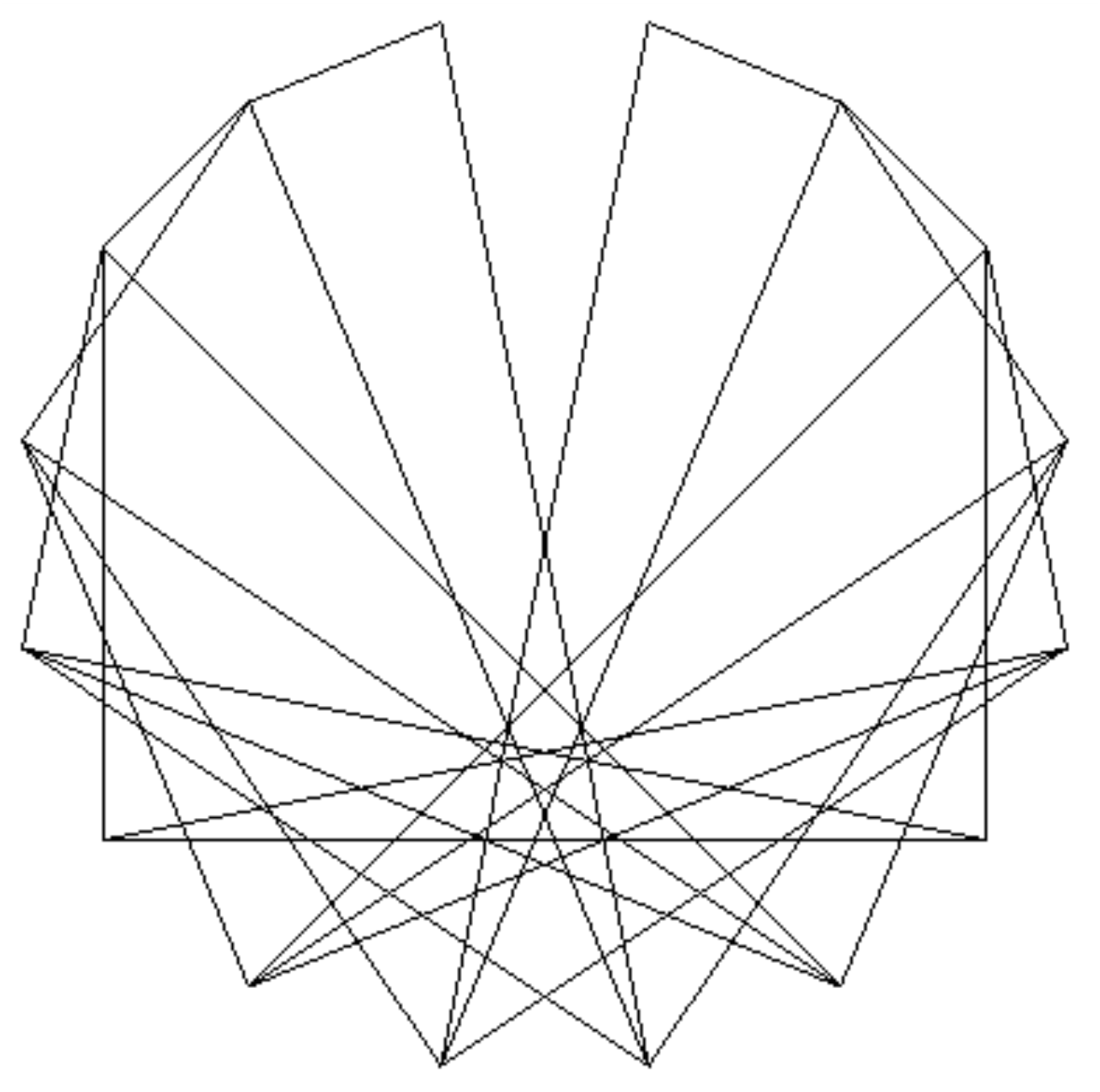}}} \hspace{1.4cm}
\subfigure[]{\scalebox{0.35}{\includegraphics{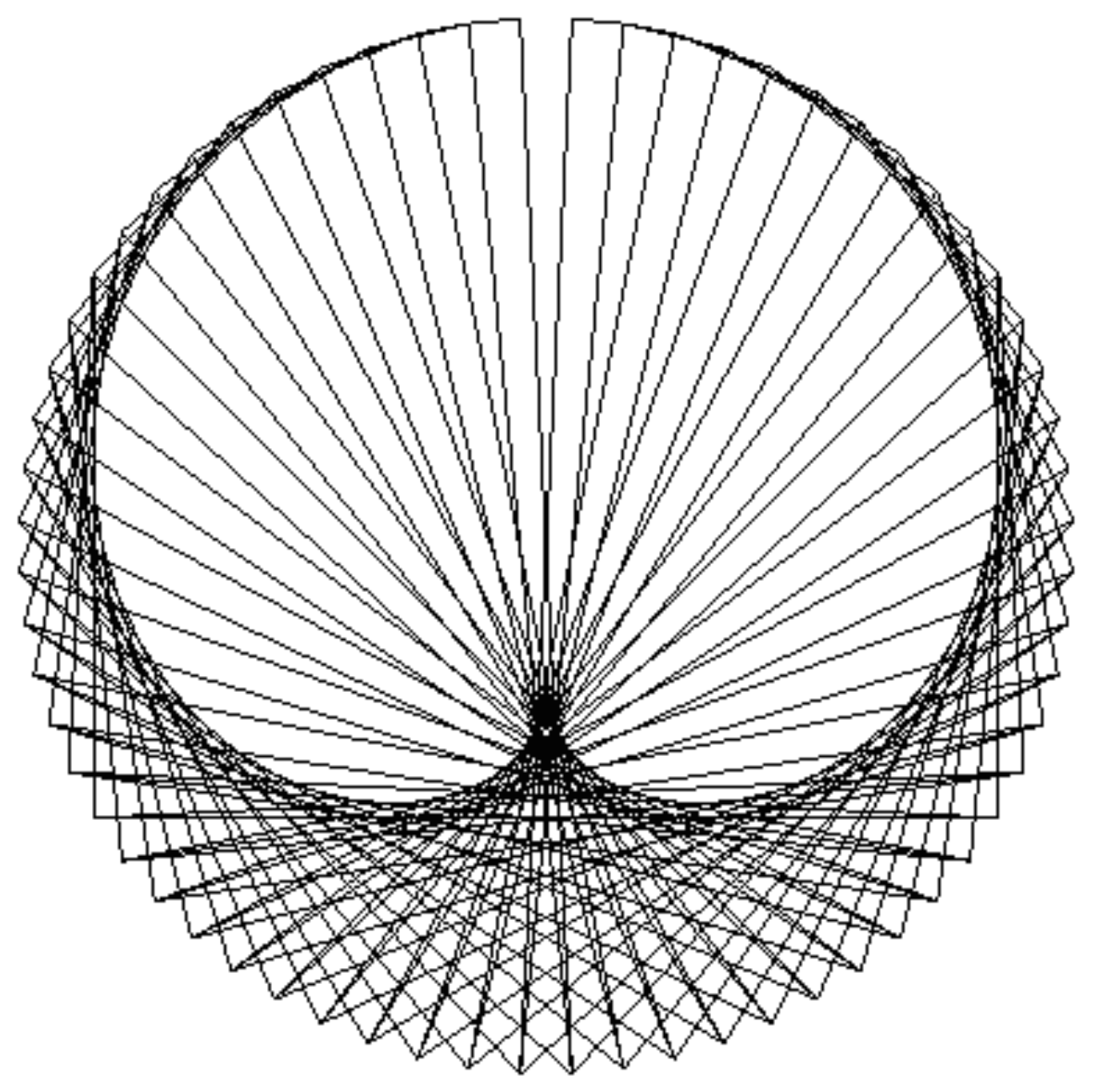}}} \hspace{1.4cm}
\subfigure[]{\scalebox{0.35}{\includegraphics{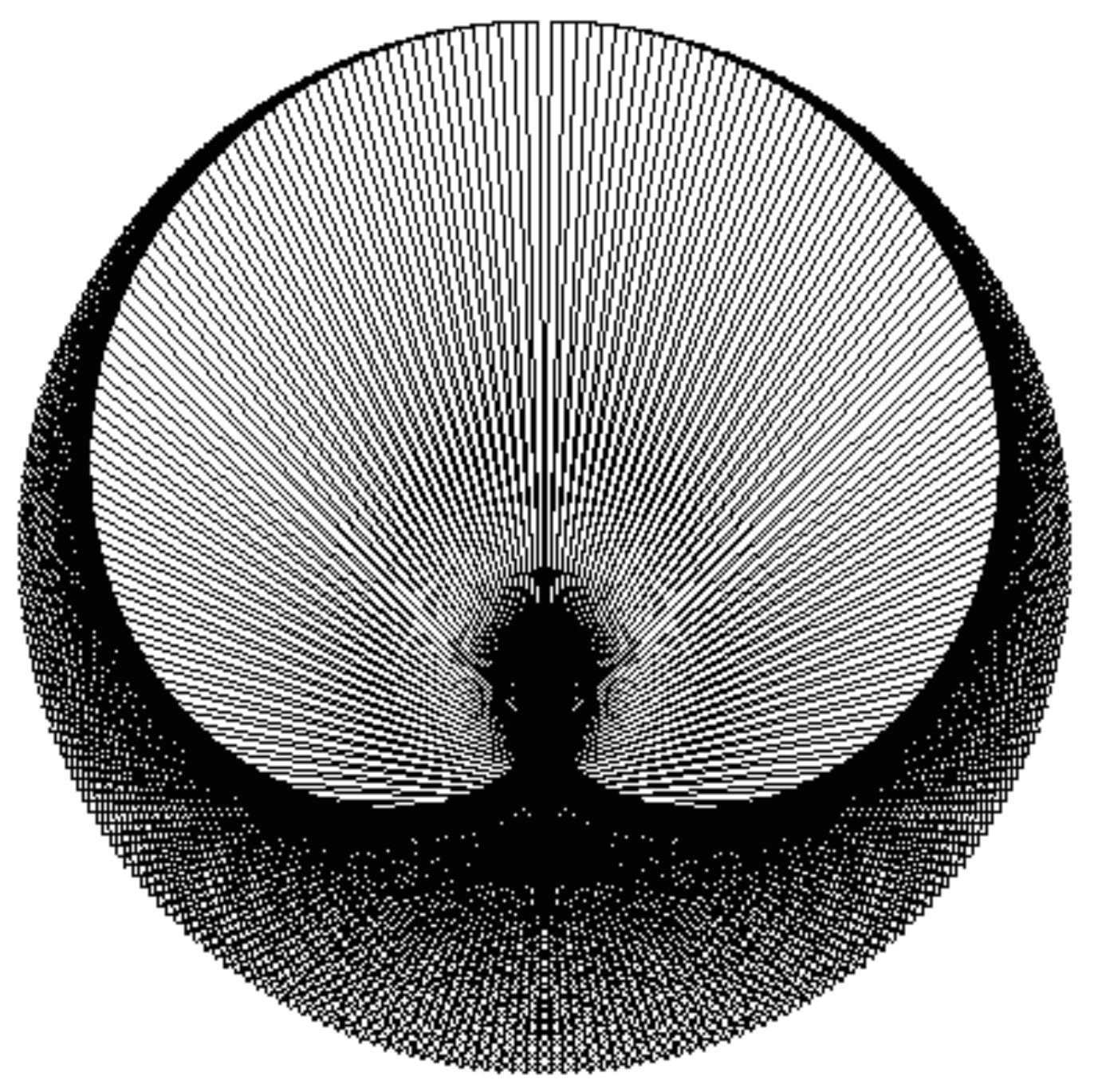}}} \hspace{1.4cm}
\subfigure[]{\scalebox{0.35}{\includegraphics{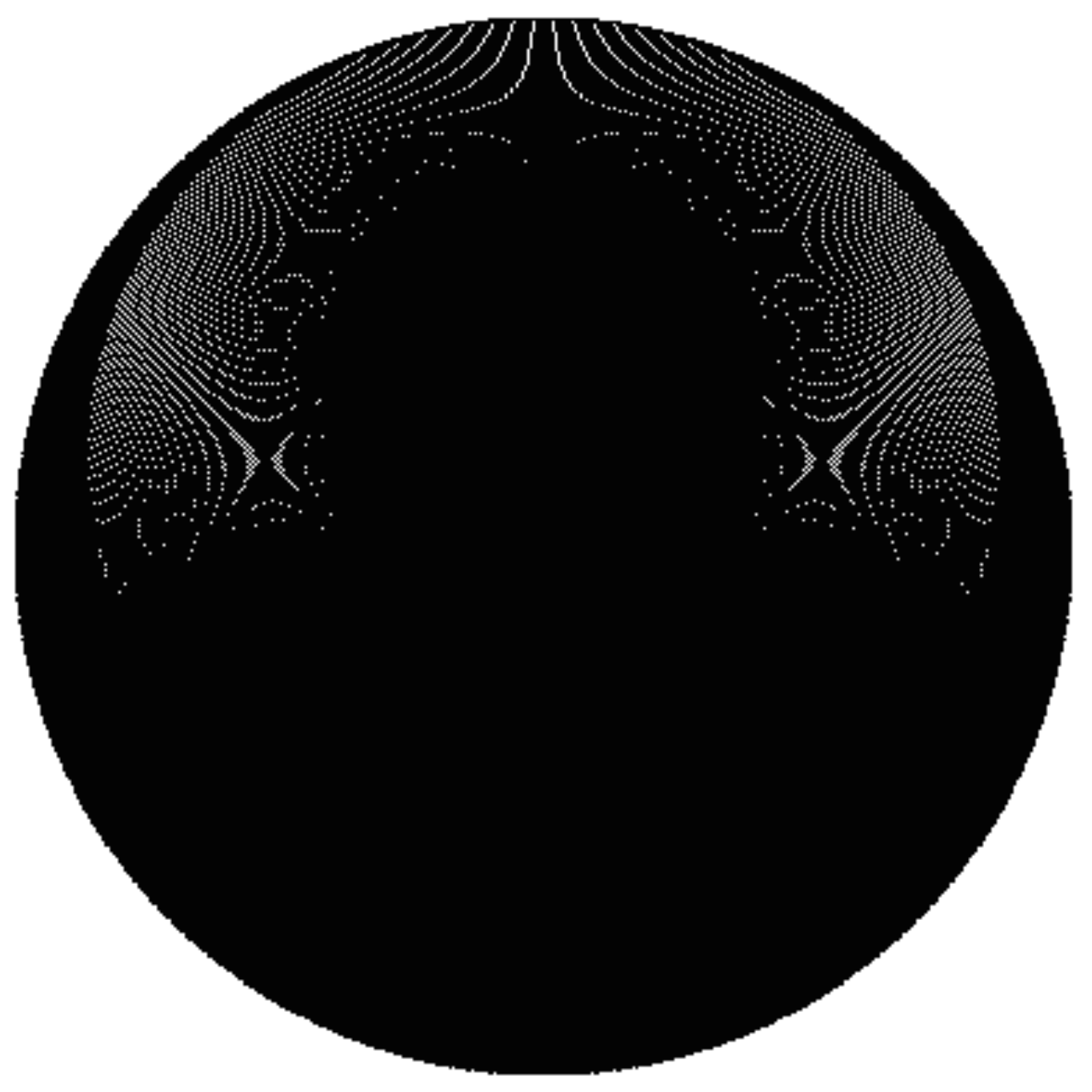}}} \hspace{1.4cm}
\subfigure[]{\scalebox{0.35}{\includegraphics{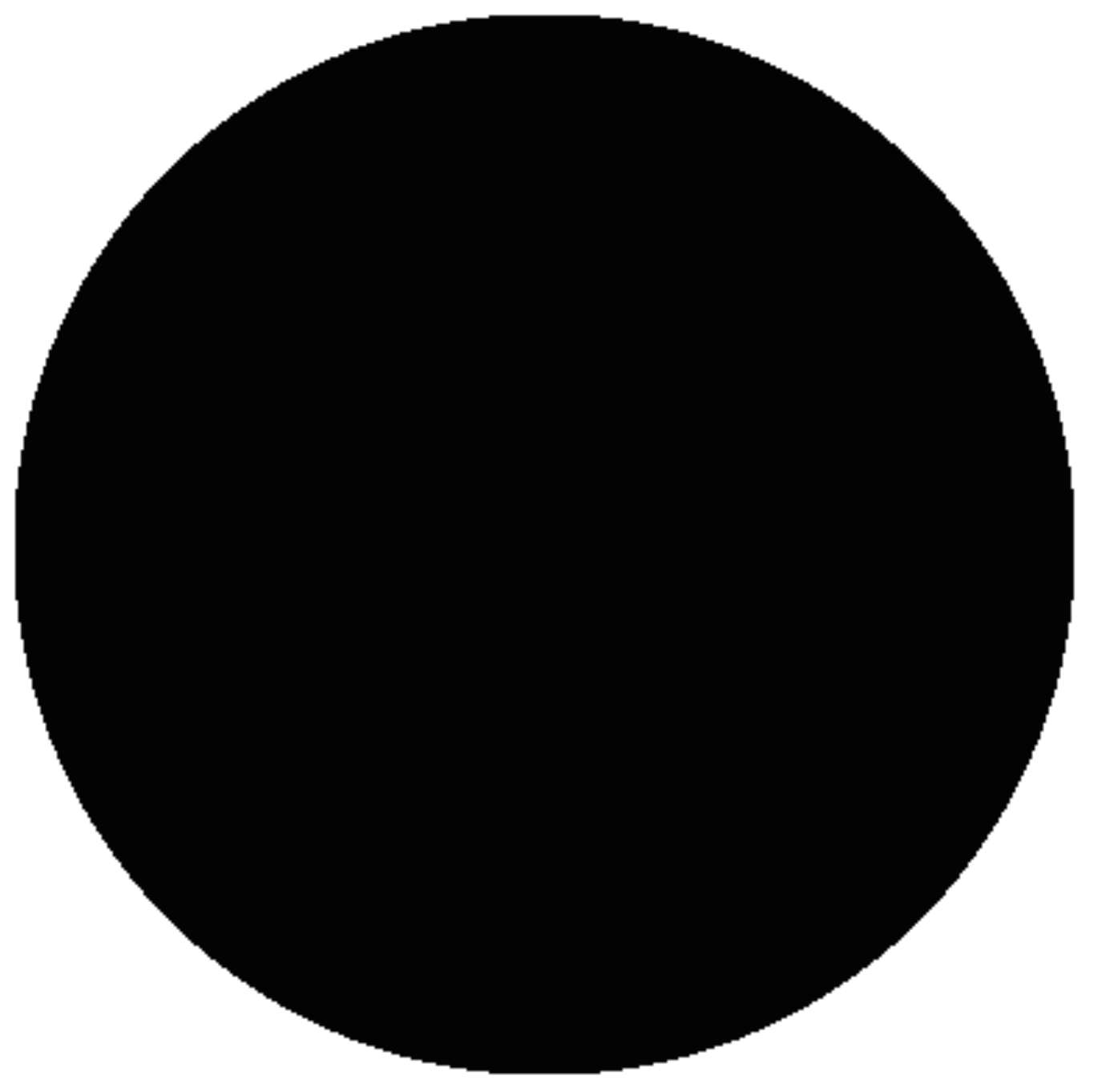}}}
\end{center}
\caption{Generic de Bruijn diagrams for ECA. Each generic diagram follows the number of partial neighbourhoods $k^{2r}$ for $r=$ (a) 1, (b) 2, (c) 3, (d) 4, (e) 5, and (f) 6.}
\label{gdb}
\end{figure}

For a one-dimensional CA of order $(k,r)$ and a finite alphabet given $\Sigma$, its de Bruijn diagram is a directed graph with $k^{2r}$ vertexes and $k^{2r+1}$ edges calculated as follows. Vertexes are labelled with elements of the alphabet of length $2r$, i.e. neighbourhood states. An edge is directed from vertex $i$ to vertex $j$, if and only if, the $2r-1$ final symbols of $i$ are the same as $2r-1$ initial symbols in $j$ forming a neighbourhood of $2r+1$ states represented by $i \diamond j$. In this case, the edge connecting $i$ to $j$ is labelled by $\varphi(i \diamond j)$ (the value of the neighbourhood defined by the local function) \cite{cavoorhees}.

Thus the de Bruijn diagram is constructed as follow:

\begin{equation}
	M_{i,j} = \left\{\begin{array}{ll}
			        1& \mbox{if } j = ki, ki+1, \ldots, ki+k-1 \mbox{ (mod } k^{2r}) \\
		           	 0 & \mbox{in other case} \\
		       \end{array}
			\right.
\label{eq-Bruijn}
\end{equation}

\begin{figure}
\includegraphics[scale=.48]{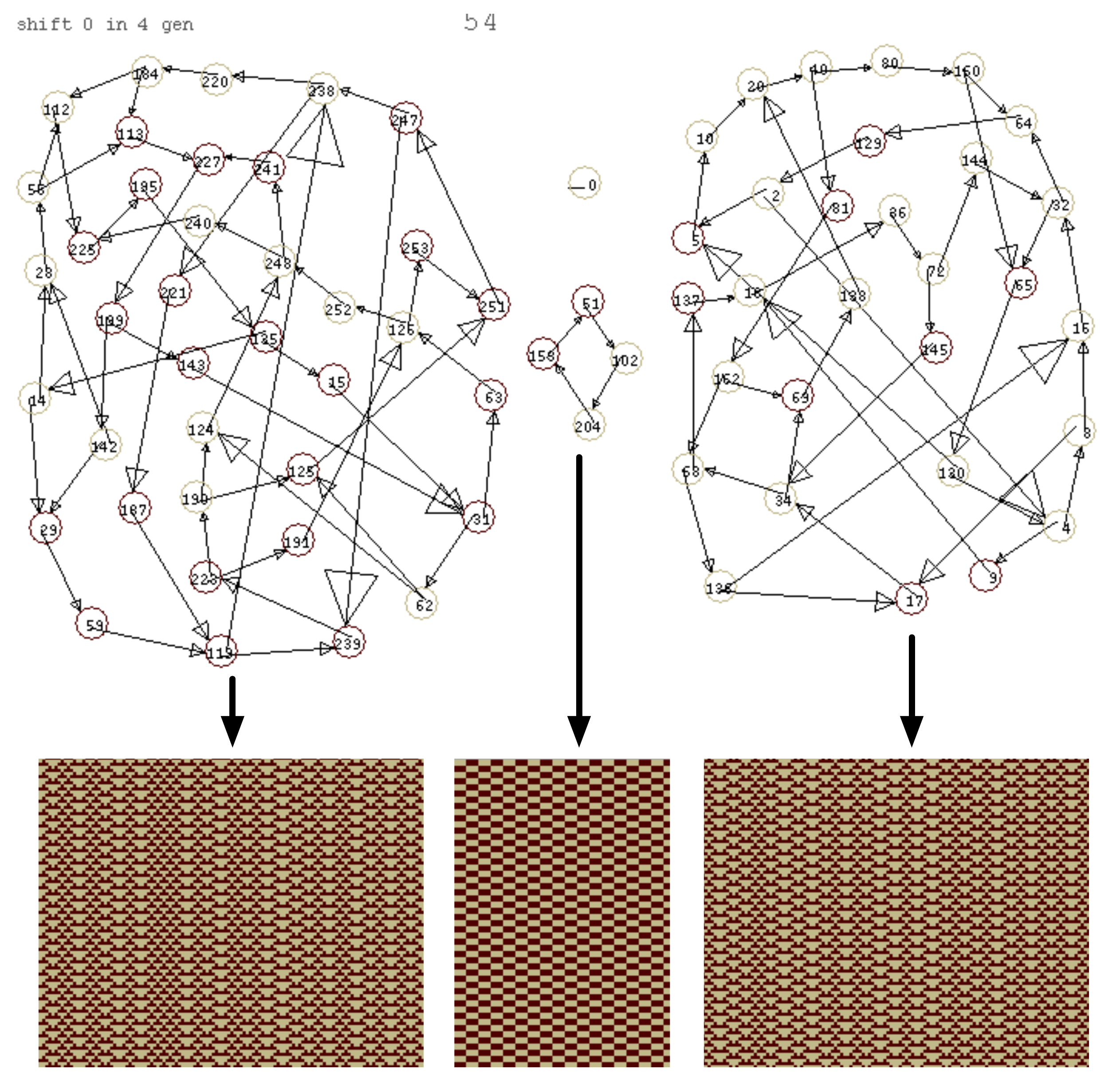}
\caption{De Bruijn diagram (0,4) calculating stationary localisations in Rule 54. A snapshot for every cycle is showed below of every diagram. This way, patterns are defined as a code since its initial condition obtained from diagram.}
\label{deBruijn0to4R54}
\end{figure}

\begin{figure}
\includegraphics[scale=.37]{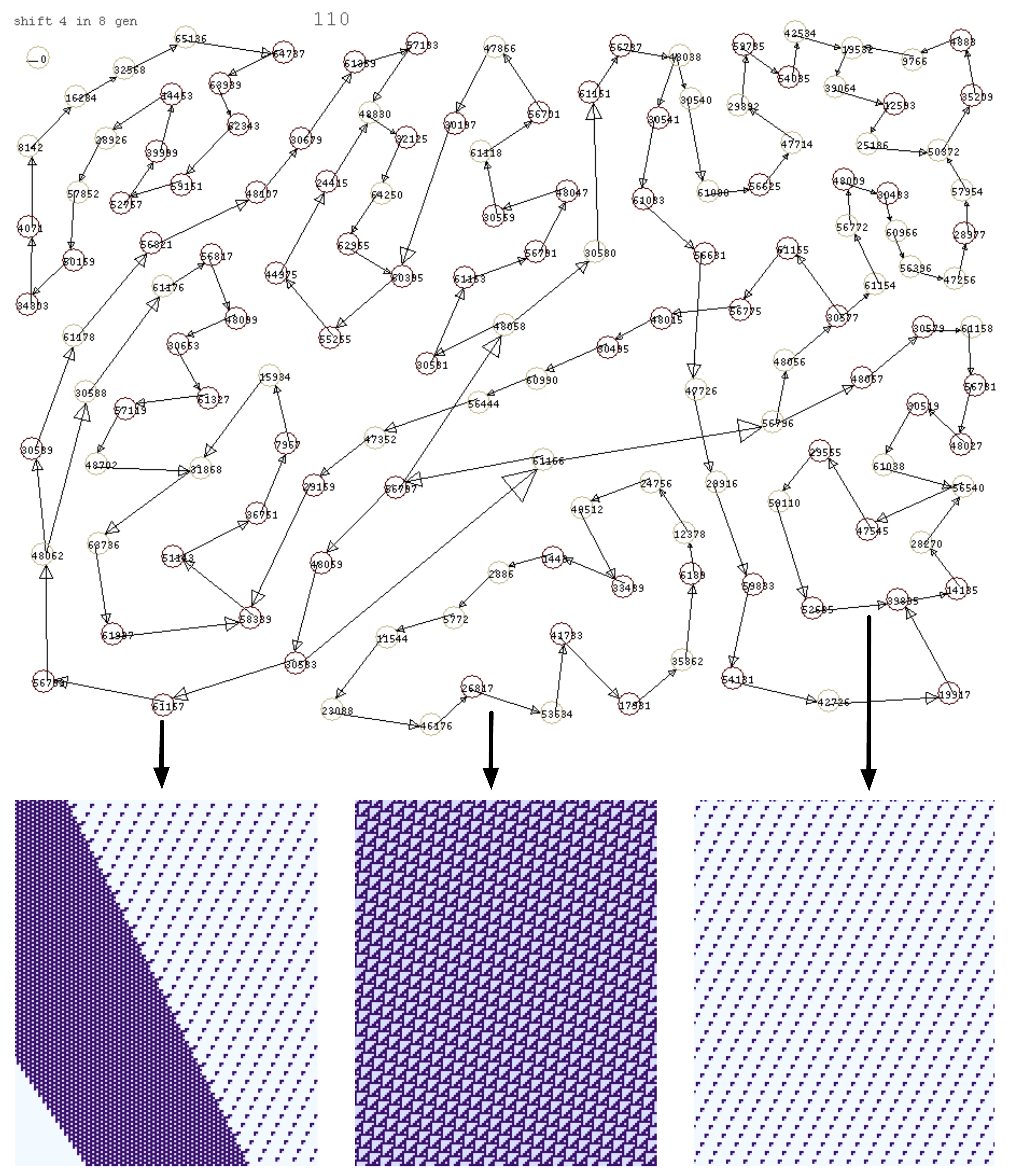}
\caption{De Bruijn diagram (4,8) calculating non-stationary localisations in Rule 110. The left evolution displays a fuse pattern produced by two mobile localisations colliding and both annihilated, the center evolution displays a periodic pattern, and the right evolution displays a mobile localisation with displacement to the left.}
\label{deBruijn8to4ECA110}
\end{figure}

For ECA the module $k^{2r}=2^{2}=4$ represents the number of vertexes in the de Bruijn diagram and $j$ takes values from $k*i=2i$ to $(k*i)+k-1=(2*i)+2-1=2i+1$. The vertexes (indexes of $M$) are labelled by fractions of neighbourhoods beginning with 00, 01, 10 and 11, the overlap determines each connection (Fig.~\ref{gdb}a). Paths in the de Bruijn diagram may represent chains, configurations, or classes of configurations in the evolution space. Also fragments of the diagram itself are useful in discovering periodic blocks of strings, pre-images, codes, and cycles~\cite{deBruijnmc}.

After the de Bruijn diagram is completed, we can calculate an extended de Bruijn diagram. An extended de Bruijn diagram takes into account more significant overlapping of neighbourhoods of length $2r$. We represent $M^{(2)}$ by indexes $i=j=2r*n$, where $n \in \mathbb Z^{+}$. The de Bruijn diagram grows exponentially, order $k^{2r^{n}}$, for each $M^{(n)}$. We can calculate generic de Bruijn diagrams arranged in a circular pattern for $r=1$ Fig.~\ref{gdb}a (basic diagram), $r=2$ Fig.~\ref{gdb}b, $r=3$ Fig.~\ref{gdb}c, $r=4$ Fig.~\ref{gdb}d, $r=5$ Fig.~\ref{gdb}e, $r=6$ Fig.~\ref{gdb}f (constructing a circle).

Generic diagrams calculate strings of different periods. These patterns are structures without displacements. The complement diagrams calculates periods plus displacements. In these diagrams we can find systematically any periodic structure, including some mobile localisations.

\begin{figure}
\includegraphics[scale=.37]{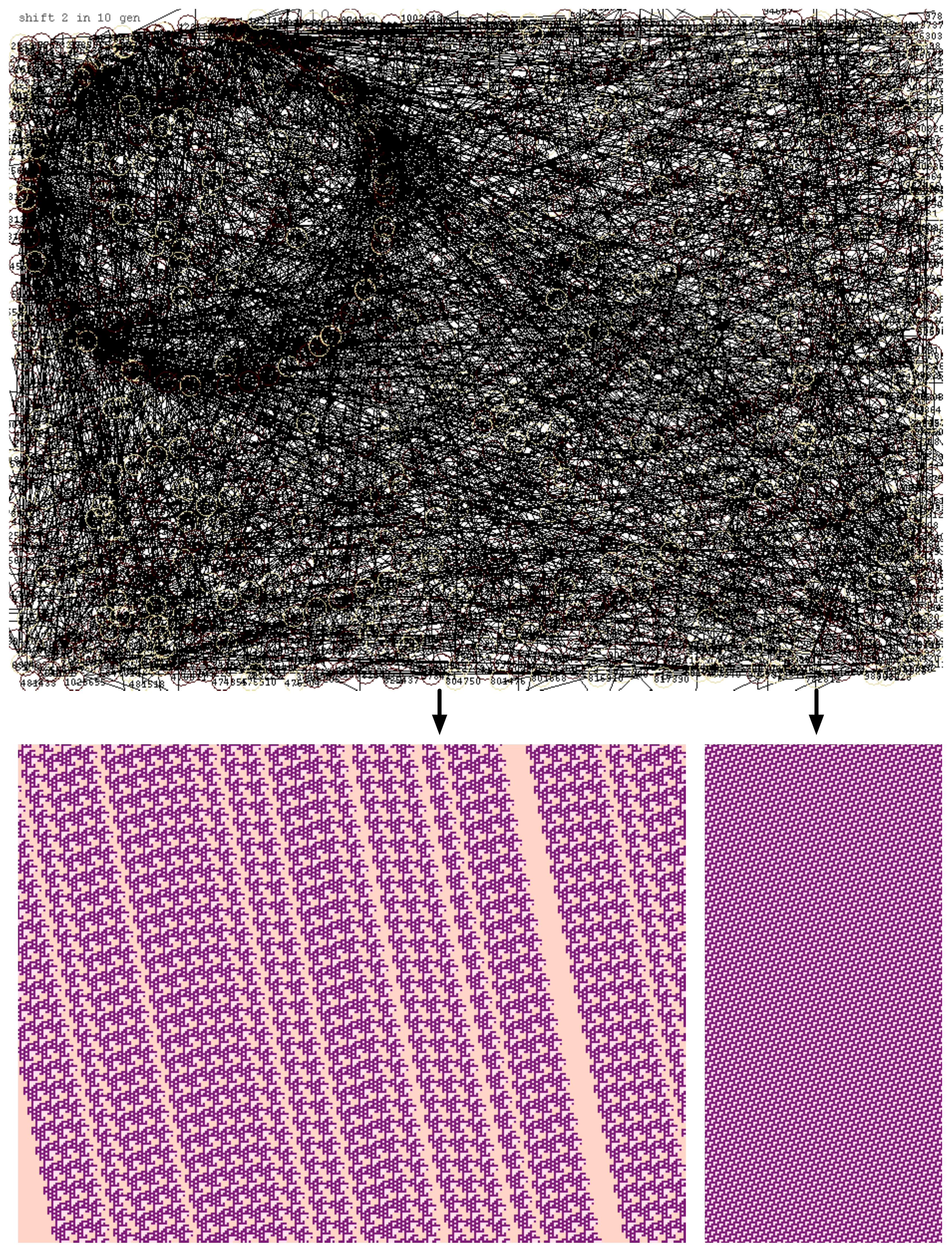}
\caption{De Bruijn diagram (2,10) calculating non-stationary localisations in Rule 110. 
First, the original diagram is calculated with 1,048,576 vertices. Below, an evolution of non-stationary localisations beginning from the vertex 652,687 (left) and a periodic background defining a small mosaic.}
\label{deBruijn10to2ECA110}
\end{figure}

For extended de Bruijn diagrams we have shift registers to the right $(+)$ or to the left $(-)$. A mobile localisation can be identified as a cycle and the localisations interaction will be a connection with other cycles. Diagram $(2,2)$ ($x$-displacements, $y$-generations), displays periodic strings moving two cells to the right in two time steps, i.e., period of a mobile localisation. This way, we can enumerate each string for every structure in this domain.

The de Bruijn diagram than can calculate stationary localisations is of order $M^{(4)}_{R54}$ because localisations have period four without displacements. These patterns can be considered also as still life configurations. Figure~\ref{deBruijn0to4R54} shows the full de Bruijn diagram (0,4) used to calculate these stationary localisations. There are four main cycles, two largest cycles represent phases of each stationary localisation plus its periodic background; and two smaller cycles characterising two different periodic patterns in Rule 54 including the stable state represented with a loop by vertex zero. Space-time configurations of ECA derived from these diagrams are illustrated on the left plate of Fig.~\ref{deBruijn0to4R54}. Position of each mobile localisation and periodic background follows arbitrarily routes into these cycles. Details on these regular expressions for Rule 54 are presented in~\cite{rule54cs}.

For Rule 110 we have calculated an extended de Bruijn diagram (4,8) that determines non-stationary localisations. Figure~\ref{deBruijn8to4ECA110} shows a diagram than initially needs 65,536 vertexes. However, we can reduce the diagram just filtering cycles, this way we have a diagram of 145 vertexes and 153 links. This way, this diagram displays a stable state represented by loop zero and cycles define a phase of a mobile localisation, i.e., a string that determines how to input a mobile localisation into its initial condition: left evolution shows mobile localisations colliding constantly, right evolution shows mobile localisations moving to the left, and center evolution shows a periodic pattern. This characteristic is very useful to control collisions of localisations in this automaton. Connections between cycles mean that you can connect several structures on a same phase. Figure~\ref{deBruijn10to2ECA110} shows the de Bruijn diagram (2,10) than calculates non-stationary localisations and periodic patterns coveting the evolution space of Rule 110. This diagram surpass a million of vertexes and shows a snapshot of a particular initial condition, coding several non-stationary localisations beginning in a diverse number of phases, copies, and intervals. Details about these regular expressions for Rule 110 can be found in \cite{reglangr110}.

De Bruijn diagrams contain all relevant information about of complex patterns emerging in cellular automata de Bruijn diagrams can proof exhaustively the number of periodic patterns that rule can yield. But they grown quickly, therefore not rarely a second algorithm must be implemented to extract useful strings from a diagram.

\section{Subsystem diagrams}

Let us focus on presenting an analytical characterisation of symbolic dynamics of mobile localisations in Rule 110. For each mobile localisation, a particular subsystem can be found through enumeration and exhaustive analysis. Directed graph theory and transition matrix provide are powerful tools for studying each sub-shift of a finite type --- such as a glider --- which is topologically mixing and possesses the positive topological entropy on this subsystem. A positive topological entropy implies the chaos in the sense of Li-Yorke \cite{LLH07, TCh07}. Devaney \cite{Dev89} and Li-Yorke types of chaos can be deduced from topological mixing. Here we describe a non-stationary localisation, other types of mobile localisations can be studied by analogy.

Particularly, a non-stationary localisation with velocity to 1/5 in Rule 110~\footnote{Gliders in Rule 110 \url{http://uncomp.uwe.ac.uk/genaro/rule110/glidersRule110.html}} can be represented as, $\Lambda_{\mathcal{A}}=\{ x\in S^{Z}| x_{[i-7,i+6]} \in \mathcal{A}, \forall i \in Z\}$ is a sub-shift of a finite type, its determinative system $\mathcal{A}= \{$9976, 3569, 7138, 14276, 12169, 7955, 15910, 15437, 14491, 12599, 8815, 1247, 2494, 4988, 8814, 1245, 2490, 4980, 9960, 3537, 7075, 14150, 11917, 7451, 14903, 13423, 10463, 4542, 9084, 1784, 2495, 4990, 9980, 3577, 7155, 14310, 12237, 8091, 16183, 15983, 15583, 14782, 13180, 3568, 7137, 14274, 12165, 7947, 15895, 15407, 14431, 12478, 8572, 760, 1521, 3042, 6084, 12168, 7953, 15907, 15431, 14479, 12574, 8764, 1144, 2289, 4578, 9156, 1929, 3859, 7718, 4989, 9979, 3575, 7150, 14300, 12217, 8051, 16103, 15823, 15263, 14142, 11900, 7416, 14833, 13282, 10180, 3977, 15436, 14489, 12595, 8806, 1228, 2457, 4914, 9828, 3273, 6547, 13094, 9805, 3227, 6455, 12911, 9439, 14301, 12219, 8055, 16111, 15839, 15295, 14206, 12028, 7672, 15345, 14306, 12228, 8073, 16147, 7139, 14279, 12174, 7965, 15930, 15477, 14571, 12758, 9132, 1880, 3761, 7522, 15044, 13705, 11027, 5670, 11341, 6299, 15931, 15479, 14574, 12764, 9144, 1904, 3809, 7618, 15236, 14089, 11795, 7206, 14413, 12443, 8503, 623, 2491, 4983, 9966, 3549, 7099, 14198, 12013, 7643, 15287, 14191, 11999, 7614, 15228, 14072, 11761$\}$. Let $x_{[i-7, i+6]}$ denote a 14-bit string $(x_{i-7}, x_{i-6}, x_{i-5}, \ldots, x_{i+5}, x_{i+6})$ over $S=\{0,1\}$ which is described by its decimal code expression, such as $9976$ refers to the string (1,0,0,1,1,0,1,1,1,1,1,0, 0,0), and $3569$ refers to the string (0,0,1,1,0,1,1,1,1,1,0,0,0,1).

\begin{figure}
\includegraphics[scale=.33]{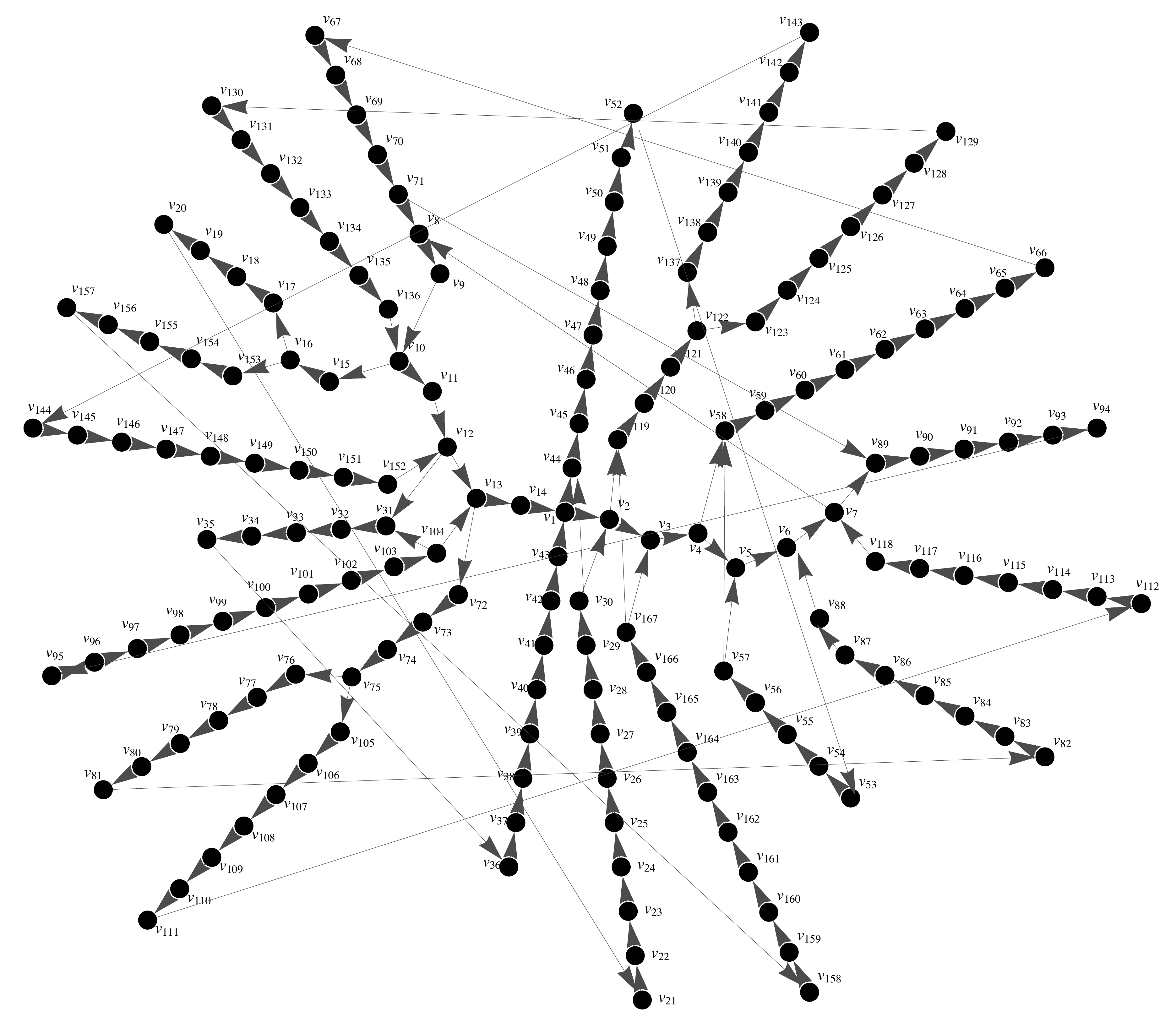}
\caption{Graph representation for the subsystem $\Lambda_{\mathcal{A}}$ of a non-stationary localisation with velocity $1/5$, it is the same mobile localisation calculated with a de Bruijn diagram showed in Fig.~\ref{deBruijn10to2ECA110}. Where each vertex stands for the element of $\mathcal{A}$ by order, i.e., $v_{1}=9976$, $v_{2}=3569$, $v_{3}=7138$, $\cdots$, $v_{166}=14072$, $v_{167}=11761$. All bi-infinite walks on the graph constitute the closed invariant subsystem $\Lambda_{\mathcal{A}}$.}
\label{D1subshift}
\end{figure}

A fundamental method for constructing finite shifts starts with a finite, directed graph and produces the collection of all bi-infinite walks (i.e., strings of edges) on the graph. The $\Lambda_{\mathcal{A}}$ can be described by a finite directed graph $G_{\mathcal{A}} = \{\mathcal{A}, E\}$, where each vertex $(V)$ is a string in $\mathcal{A}$. Each edge $e\in E(G)$ starts at a string denoted by $a=(a_{0}, a_{1}, \cdots, a_{13})\in \mathcal{A}$ and terminates at the string $b=(b_{0}, b_{1}, \cdots, b_{13})\in V(G)$ if and only if $a_{k} = b_{k-1}, 1\leq k \leq 13$. One can represent each element of $\Lambda_{\mathcal{A}}$ as a certain path on the graph $G_{\mathcal{A}}$. The entire bi-infinite walks on the graph constitute the closed invariant subsystem $\Lambda_{\mathcal{A}}$. The finite directed graph $\Lambda_{\mathcal{A}}$, is shown in Fig.~\ref{D1subshift}. A finite path $P=v_{1}\rightarrow v_{2}\rightarrow \cdots \rightarrow  v_{m}$ on a graph $G$ is a finite string of vertexes $v_{i}$ from $G$. The length of $P$ is $|P|=m$. A cycle is a path that starts and terminates at the same vertex. A graph $G$ is irreducible if for each ordered pair of vertexes $I$ and $J$ there is a path in $G$ starting at $I$ and terminating at $J$. Each element of $\Lambda_{\mathcal{A}}$ is a certain path on the graph $G_{\mathcal{A}}$.

Let $\widehat{S}=\{r_{0},r_{1},\cdots,r_{165},r_{166}\}$ be a new symbolic set, where $r_{i},i=0,\cdots,166$, stands for the element of $\mathcal{A}$ respectively. Then, one can construct a new symbolic space $\widehat{S}^{Z}$ on $\widehat{S}$, where $\mathcal{B}=\{(rr')|r=(b_{0}b_{1}\cdots b_{13}), r'=(b_{0}'b_{1}'\cdots b_{13}')\in \widehat{S},~\forall~2 \leq j\leq 13 \ s. t.\  b_{j}=b_{j-1}'\}$. The two-order sub-shift $\Lambda_{\mathcal{B}}$ of $\sigma$ is defined as $\Lambda_{\mathcal{B}}=\{r=(\cdots,r_{-1},{r}_{0}^{\ast},r_{1},\cdots )\in \widehat{S}^{Z}|r_{i}\in \widehat{S}, (r_{i},r_{i+1})\prec \mathcal{B},\forall i\in Z\}$. One can calculate the transition matrix $\mathcal{Y}$ of the sub-shift $\Lambda_{\mathcal{B}}$ with $\mathcal{Y}_{ij}=1$, if $(i,j)\prec\Lambda_{\mathcal{B}}$; otherwise $\mathcal{Y}_{ij}=0$. We call the matrix $\mathcal{Y}$ positive if all of its entries are non-negative; irreducible if $\forall i,j$, there exist $n$ such that $\mathcal{Y}_{ij}^{n}>0$; aperiodic if there exists $N$, such that $\mathcal{Y}_{ij}^{n}>0,n>N,\forall~i,j$. Then, $\Lambda_{\mathcal{B}}$ is topologically mixing if and only if $\mathcal{Y}$ is irreducible and aperiodic. Thus, the topologically conjugate relationship between $(\Lambda_{\mathcal{A}},\sigma)$ and a two-order sub-shift of finite type $(\Lambda_{\mathcal{B}},\sigma)$ can be established. In addition, the transition matrix $\mathcal{Y}$ is relatively large (the order of $D$ is 167). Therefore, we only list the indexes $(i,j)$ of nonzero elements. $\mathcal{Y}=\{$$(1, 2)$, $(1, 44)$, $(2, 3)$, $(2, 119)$, $(3, 4)$, $(4, 5)$, $(4, 58)$, $(5, 6)$, $(6, 7)$, $(7, 8)$, $(7, 89)$, $(8, 9)$, $(9, 10)$, $(10, 11)$, $(10, 15)$, $(11, 12)$, $(12, 13)$, $(12, 31)$, $(13, 14)$, $(13, 72)$, $(14, 1)$, $(15, 16)$, $(16, 17)$, $(16, 153)$, $(17, 18)$, $(18, 19)$, $(19, 20)$, $(20, 21)$, $(21, 22)$, $(22, 23)$, $(23, 24)$, $(24, 25)$, $(25, 26)$, $(26, 27)$, $(27, 28)$, $(28, 29)$, $(29, 30)$, $(30, 2)$, $(30, 44)$, $(31, 32)$, $(32, 33)$, $(33, 34)$, $(34, 35)$, $(35, 36)$, $(36, 37)$, $(37, 38)$, $(38, 39)$, $(39, 40)$, $(40, 41)$, $(41, 42)$, $(42, 43)$, $(43, 1)$, $(44, 45)$, $(45, 46)$, $(46, 47)$, $(47, 48)$, $(48, 49)$, $(49, 50)$, $(50, 51)$, $(51, 52)$, $(52, 53)$, $(53, 54)$, $(54, 55)$, $(55, 56)$, $(56, 57)$, $(57, 5)$, $(57, 58)$, $(58, 59)$, $(59, 60)$, $(60, 61)$, $(61, 62)$, $(62, 63)$, $(63, 64)$, $(64, 65)$, $(65, 66)$, $(66, 67)$, $(67, 68)$, $(68, 69)$, $(69, 70)$, $(70, 71)$, $(71, 8)$, $(71, 89)$, $(72, 73)$, $(73, 74)$, $(74, 75)$, $(75, 76)$, $(75, 105)$, $(76, 77)$, $(77, 78)$, $(78, 79)$, $(79, 80)$, $(80, 81)$, $(81, 82)$, $(82, 83)$, $(83, 84)$, $(84, 85)$, $(85, 86)$, $(86, 87)$, $(87, 88)$, $(88, 6)$, $(89, 90)$, $(90, 91)$, $(91, 92)$, $(92, 93)$, $(93, 94)$, $(94, 95)$, $(95, 96)$, $(96, 97)$, $(97, 98)$, $(98, 99)$, $(99, 100)$, $(100, 101)$, $(101, 102)$, $(102, 103)$, $(103, 104)$, $(104, 13)$, $(104, 31)$, $(105, 106)$, $(106, 107)$, $(107, 108)$, $(108, 109)$, $(109, 110)$, \\$(110, 111)$, $(111, 112)$, $(112, 113)$, $(113, 114)$, $(114, 115)$, $(115, 116)$, $(116, 117)$, $(117, 118)$, $(118, 7)$, $(119, 120)$, $(120, 121)$, $(121, 122)$, $(122, 123)$, $(122, 137)$, \\$(123, 124)$, $(124, 125)$, $(125, 126)$, $(126, 127)$, $(127, 128)$, $(128, 129)$, $(129, 130)$, $(130, 131)$, $(131, 132)$, $(132, 133)$, $(133, 134)$, $(134, 135)$, $(135, 136)$, $(136, 10)$, $(137, 138)$, $(138, 139)$, $(139, 140)$, $(140, 141)$, $(141, 142)$, $(142, 143)$, $(143, 144)$, $(144, 145)$, $(145, 146)$, $(146, 147)$, $(147, 148)$, $(148, 149)$, $(149, 150)$, $(150, 151)$, $(151, 152)$, $(152, 12)$, $(153, 154)$, $(154, 155)$, $(155, 156)$, $(156, 157)$, $(157, 158)$, $(158, 159)$, $(159, 160)$, $(160, 161)$, $(161, 162)$, $(162, 163)$, $(163, 164)$, $(164, 165)$, $(165, 166)$, $(166, 167)$, $(167, 3)$, $(167, 119)\}$.

Denote the elements of $\mathcal{Y}^{n}$ as $\mathcal{Y}_{i,j}^{n}$ ,$1\leq i,j \leq 167$. $\mathcal{Y}_{i,j}^{n}$ indicates the number of the whole paths from $i$-th vertex to $j$-th vertex whose length is $n$. Thus, $\mathcal{Y}_{i,i}^{n}$ is the number of all cycles of $i$-th vertex with the length $n$. As all $\mathcal{Y}_{i,i}^{n}$, $1\leq i \leq 167$ are positive for $n=39$, it is easy to verify that each vertex possesses a particular cycle. Hence, the non-wandering set of subsystem $\Omega(\Lambda_{\mathcal{A}})=\Lambda_{\mathcal{A}}$.

The topological dynamics of $\Lambda_{\mathcal{A}}$ is  determined by the properties of its transition matrix $\mathcal{Y}$. The characteristic equation of $\mathcal{Y}$ is $-\lambda^{73}-\lambda^{84}-3\lambda^{95}+5\lambda^{106}+\lambda^{109}+5\lambda^{120}+15\lambda^{131}+10\lambda^{142}+
\lambda^{153}-\lambda^{167}=0$. The spectral radius $\rho(\mathcal{Y})$ is the maximum positive real root $\lambda^{\ast}$ of characteristic equation. We have $ent(\Lambda_{\mathcal{A}})=log(\rho(\mathcal{Y}))=log(1.12316)=0.116146$. Recall that two topologically conjugate systems have the same topological entropy and the topological entropy of $\sigma_{L}$ on $\Lambda_{\mathcal{A}}$ equals $log\rho(\mathcal{Y})$. The topological entropy $ent(\Lambda_{\mathcal{A}})$ is positive.

The matrix $(\mathcal{Y}+\mathcal{I})^{n}$ is positive for $n\geq 47$, where $\mathcal{I}$ is a $167\times167$ identity matrix. $\mathcal{Y}+\mathcal{I}$ is aperiodic, and $\mathcal{Y}$ is irreducible. $\Lambda_{\mathcal{A}}$ is topologically transitive because the transition matrix $\mathcal{Y}$ is irreducible.

A two-order sub-shift of a finite type $\Lambda_{\mathcal{A}}$ is topologically mixing if and only if its transition matrix is irreducible and aperiodic. Meanwhile, it is easy to verify that $\mathcal{Y}^{n}$ is positive for $n\geq 225$, which implies that $\mathcal{Y}$ is irreducible and aperiodic. The $\Lambda_{\mathcal{A}}$ is topologically mixing.

In conclusion, the above discussions are summarised as the subsystem $\Lambda_{\mathcal{A}}$ is chaotic in the sense of both Li-Yorke and Devaney.

\section{Basins of attraction}

Basins of attraction or cycle diagrams calculate attractors in a dynamical system, as was extensively studied by Andrew Wuensche in CA and random Boolean networks \cite{kn:WL92, ddlabbook, mcbook}.

Given a sequence of cells $x_i$ we define a configuration $c$ of the system. An evolution is represented by a sequence of configurations $\{c_0, c_1, c_2, \ldots, c_{m-1}\}$ given by the global mapping, 

\begin{equation}
\Phi:\Sigma^n \rightarrow \Sigma^n
\end{equation}

\noindent and the global relation is given for the next function between configurations,

\begin{equation}
\Phi(c^t) \rightarrow c^{t+1}.
\end{equation}

\begin{figure}[th]
\includegraphics[scale=.353]{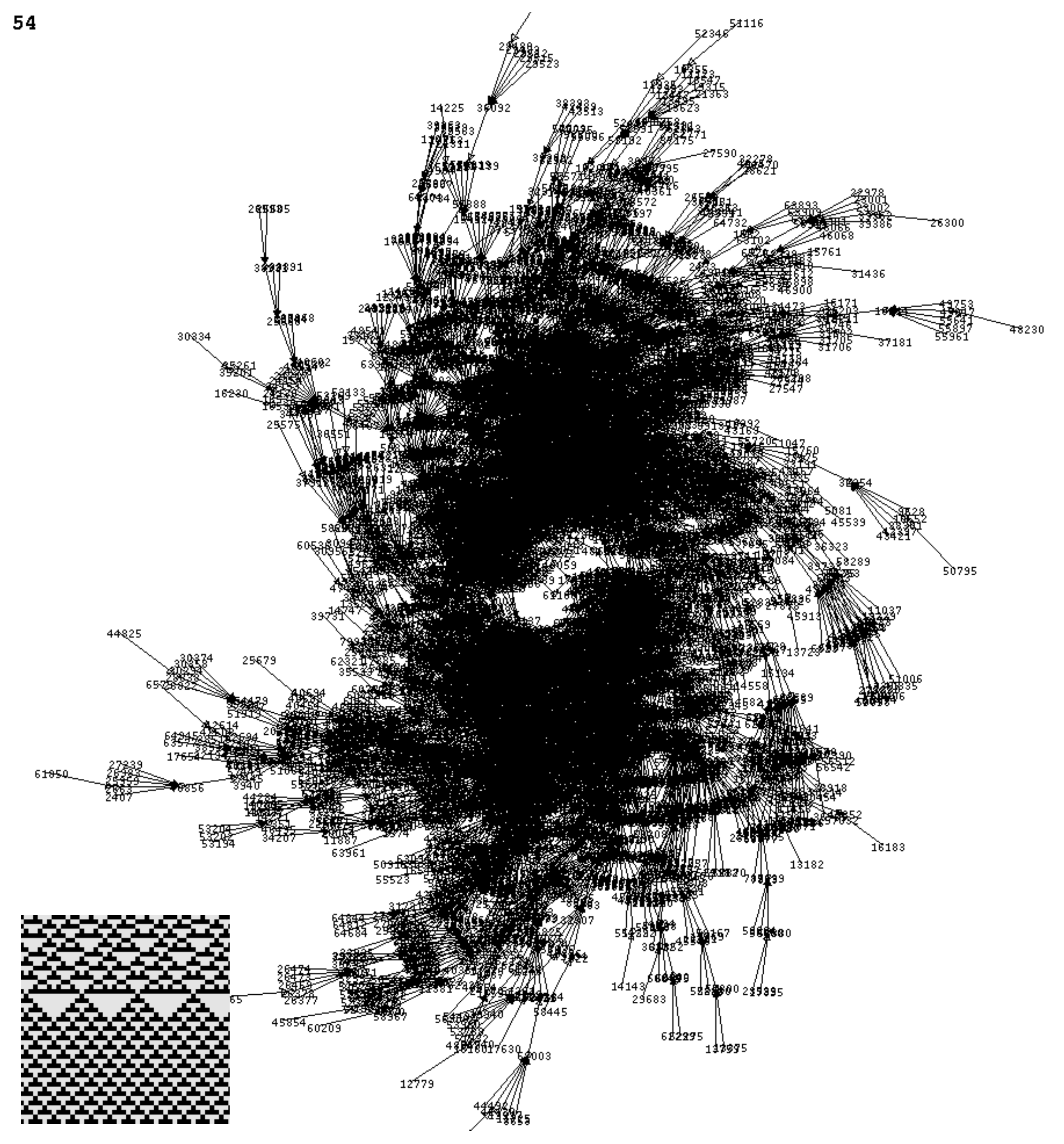}
\caption{Cycle diagram length 16, period four, with a mass of 6,432 configurations for Rule 54.}
\label{attractor16R54cycle4}
\end{figure}

An attractor is represented as a number of $c_i$ states, these states are connected in cycles in periods of global states.

To obtain cycles for a given automaton we enumerate all the rings of the desired length, and follow up their evolution. In doing so task, various shortcuts can be taken, such as generating the configurations in some order so that only a single cell changes state from one to the next, this way a number of configurations can be compacted to avoid calculating again the same string. Still life configurations and small oscillators can be detected very quickly in this way. Comparison of successive generations means that whenever the new generation is smaller, it has been already examined and there is no need for further exploration \cite{mcbook}.

The number of global states $c$ is defined by the length of the string $n^m$. The structure of an attractor is given in three parts. Leaves represent Garden of Eden configurations for those global states, that means that these states have no ancestors. Branches are configurations that have at least an ancestor and just one successor. Height in branches determines the number of generations necessary before to reach the attractor. So, the attractor is the final state of a string of length $n$. Numbers labelling vertices's represent the decimal value for the string in study.

\begin{figure}[th]
\includegraphics[scale=.353]{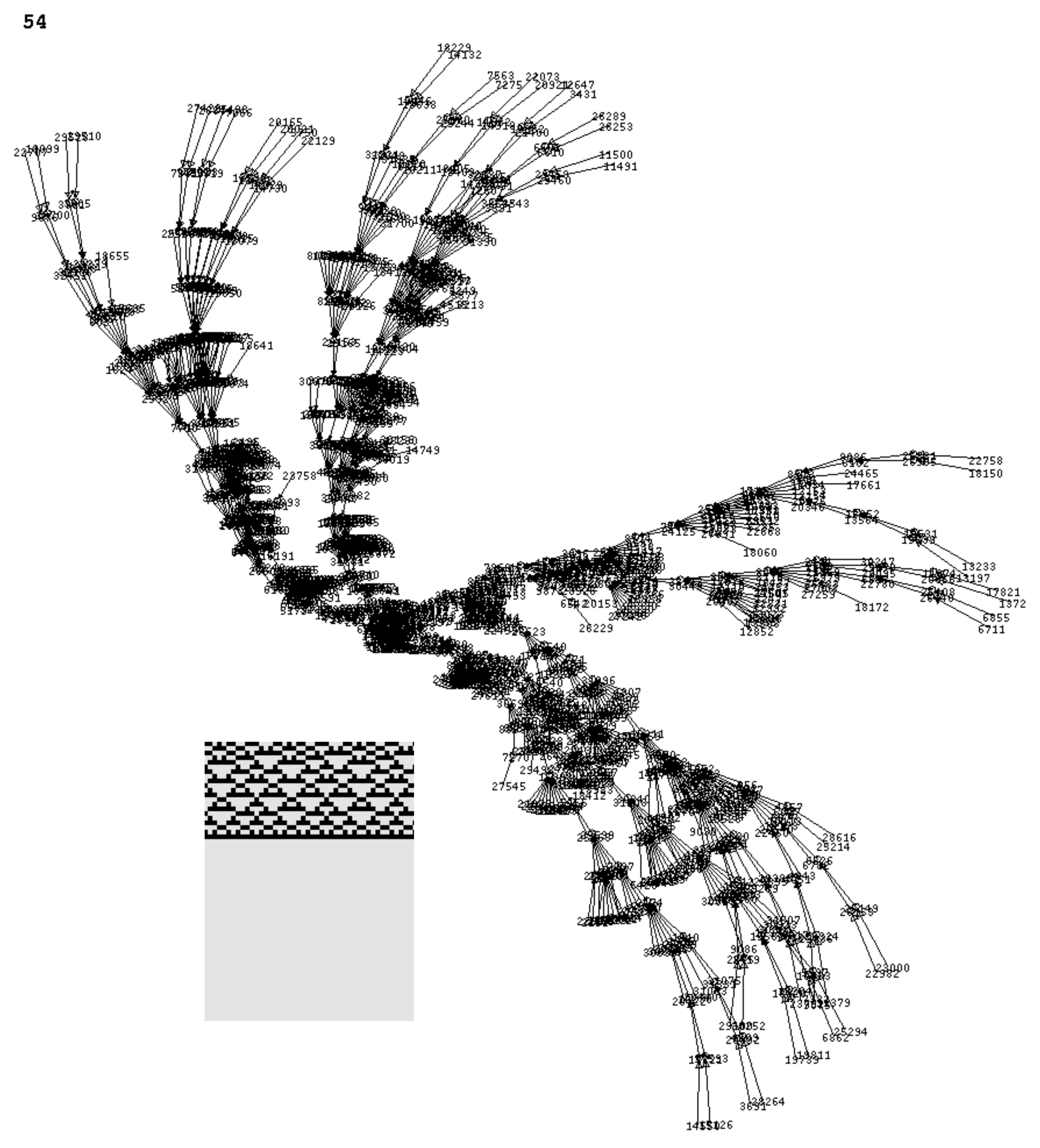}
\caption{Cycle diagram length 15, period 1, with a mass of 1,583 configurations for Rule 54.}
\label{attractor15R54cycle1}
\end{figure}

Some properties are immediate. An attractor period one (loop) determines an evolution dominated just for one state of the alphabet. Basins constructed by cycles for any $n$, they represent exactly reversible CA~\cite{reversiblemc, reversibleca}, where you can move back in the history of the system from any configuration, and therefore any configuration has one ancestor.

Generally a basin can recognise CA also with chaotic or complex behaviour using prior results on attractors \cite{kn:WL92}. This way, Wuensche discovered that Wolfram's classes can be represented as a {\it basin classification}. Talking about complex rules, a basin will be defined as:

\begin{center}
{\it Class IV: moderate transients, moderate-length periodic attractors, moderate in-degree, very moderate leaf density (possibly complex dynamics).}
\end{center}

Figure~\ref{attractor16R54cycle4} shows an attractor from a configuration of length 16 for Rule 54. This basin is made of 6,432 configurations with an attractor of four configurations. Starting from any configuration in the tree the final configuration is defined by an attractor of period four at the centre. This attractor represents concatenations of configurations: 1000, 1101, 0010, and 0111. A small evolution (left-down) shows its dynamics beginning from the state 48,230 and evolving during 20 generations to reach its final dynamics.

\begin{figure}[th]
\includegraphics[scale=.165]{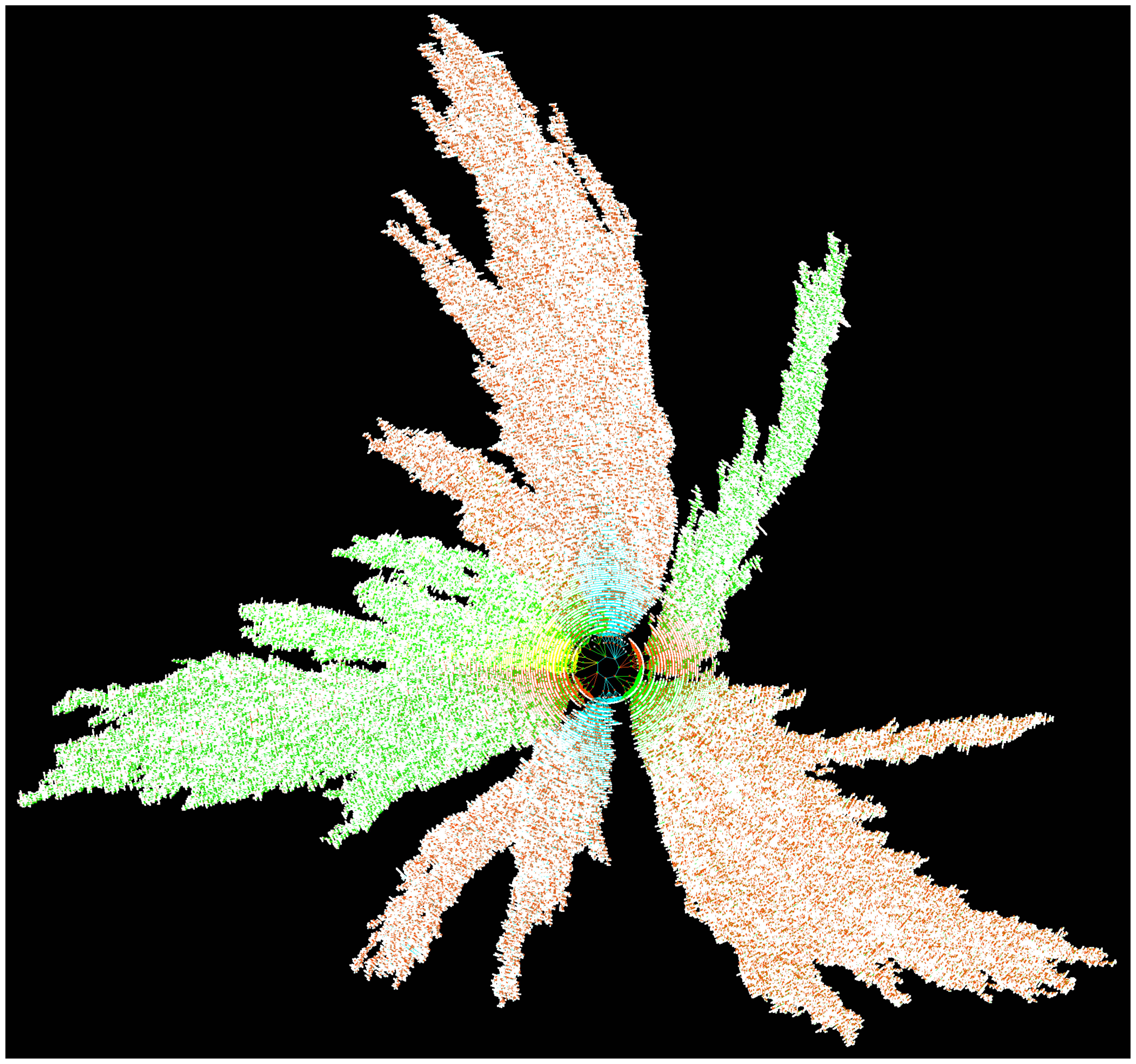}
\caption{Cycle diagram length 31, period 7, with a mass of 54,875,513 configurations for Rule 110.}
\label{basinR110size31p7} 
\end{figure}

Figure~\ref{attractor15R54cycle1} displays an attractor from a configuration of length 15 for Rule 54. This basin shows a configuration that eventually will reach to one stable state; the attractor has period one. The tree has a mass of 1,583 configurations and a height of 20 generations.

Figure~\ref{basinR110size31p7} displays a large attractor from a configuration of length 31 for Rule 110. This attractor included configurations with stationary localisations which can be referred as a structure than does not move in the evolution space. It is known as a still life configuration, this name comes originally from the famous two-dimensional binary CA Conway's the Game of Life \cite{glca}. These mobile localisations are produced from large and dense ramifications. 

\begin{figure}[th]
\begin{center}
\subfigure[]{\scalebox{0.275}{\includegraphics{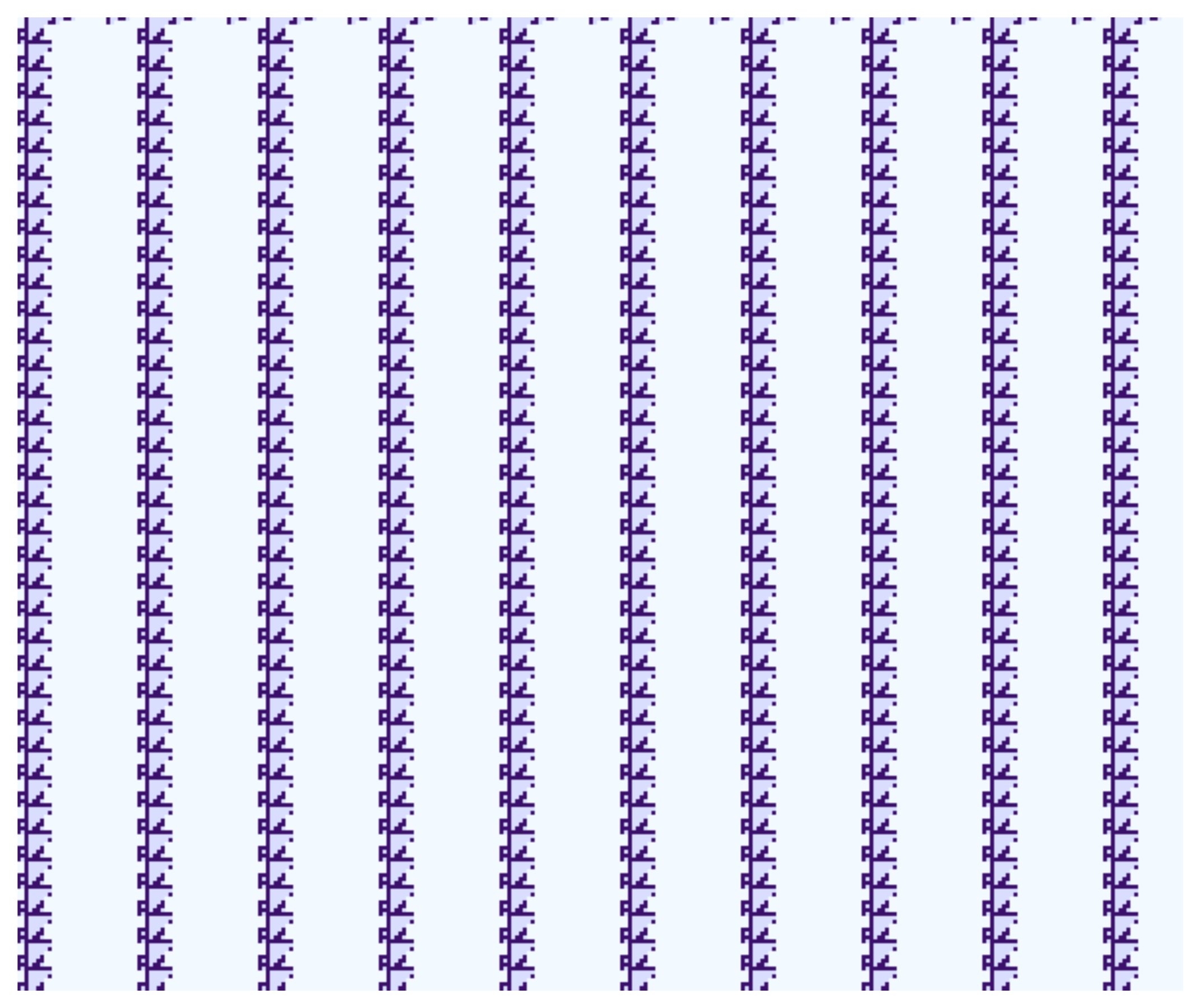}}} 
\subfigure[]{\scalebox{0.275}{\includegraphics{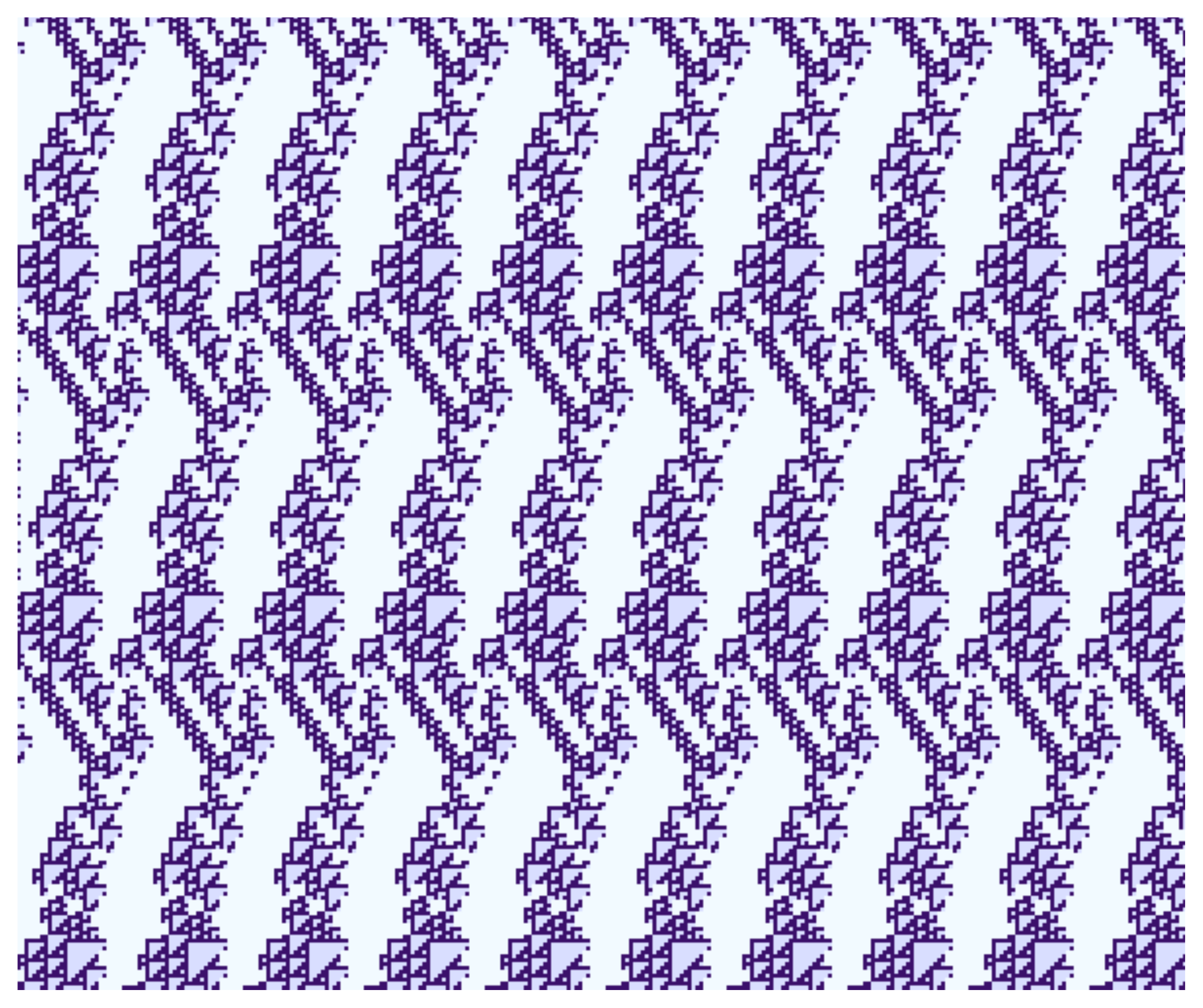}}}
\end{center}
\caption{Patterns formation from two configurations of length 31 (leaves) in Rule 110 calculated in Fig.~\ref{basinR110size31p7}. 
(a) stationary localisations. 
(b) non-trivial mobile localisations colliding. A filter is selected to a better view of localisations.}
\label{basinR110size31p7Evol}
\end{figure}

The history of a branch evolve quickly in chaotic regions with collisions of mobile localisations before it reaches the attractor. In this case, the attractor is a cycle period 7 on rings of 31 cells, with a mass of 54,875,513 configurations to construct this basin, on a space of $k^{31}$ deriving just 6,326 basins. The attractor is defined for the next configurations: 0110111111110001001101111100010 $\rightarrow$ 1111100000010011011111000100110 $\rightarrow$ 1000100000110111110001001101111 $\rightarrow$ 1001100001111100010011011111000 $\rightarrow$ 1011100011000100110111110001001 $\rightarrow$ 1110100111001101111100010011011 $\rightarrow$ 0011101101011111000100110111110.
 

Some attractors of these lengths present periodic structures. The attractor can be reached from many leaves, two of them are:

\begin{itemize}
\item 1111100001111100101000100000011 (branch 2,084,458,755),
\item 1111100011101110001110111110100 (branch 2,088,181,236).
\end{itemize}

Its dynamics display fragments of stationary and non-stationary localisations emerging during 500 and 800 generations. Starting from these configurations we can see stationary localisations or collisions of them before to reach the attractor. Figure~\ref{basinR110size31p7Evol} displays patterns from the previous strings of length 31. Concatenation of branch 2,084,458,755 yield stationary localisations (Fig.~\ref{basinR110size31p7Evol}a) and branch 2,088,181,236 yield a non-trivial synchronisation of mobile localisations colliding (Fig.~\ref{basinR110size31p7Evol}b).

By calculating large attractors, we can discover landscapes of complexity in basins featured with non-symmetric, high, and dense ramifications: these kind of ramifications are indicators of `unpredictable' behaviour on most large configurations. Frequently chaotic rules tend to have symmetric basins.

\section{Jump-graphs}

Basins of attraction can be constructed into a meta network, called the `jump-graph' by Wuensche~\cite{ddlabbook}. Jump-graphs determine the next level of CA complexity by showing a probability to jump to another attractor given a mutation on the same domain of strings.

\begin{figure}
\includegraphics[scale=.105]{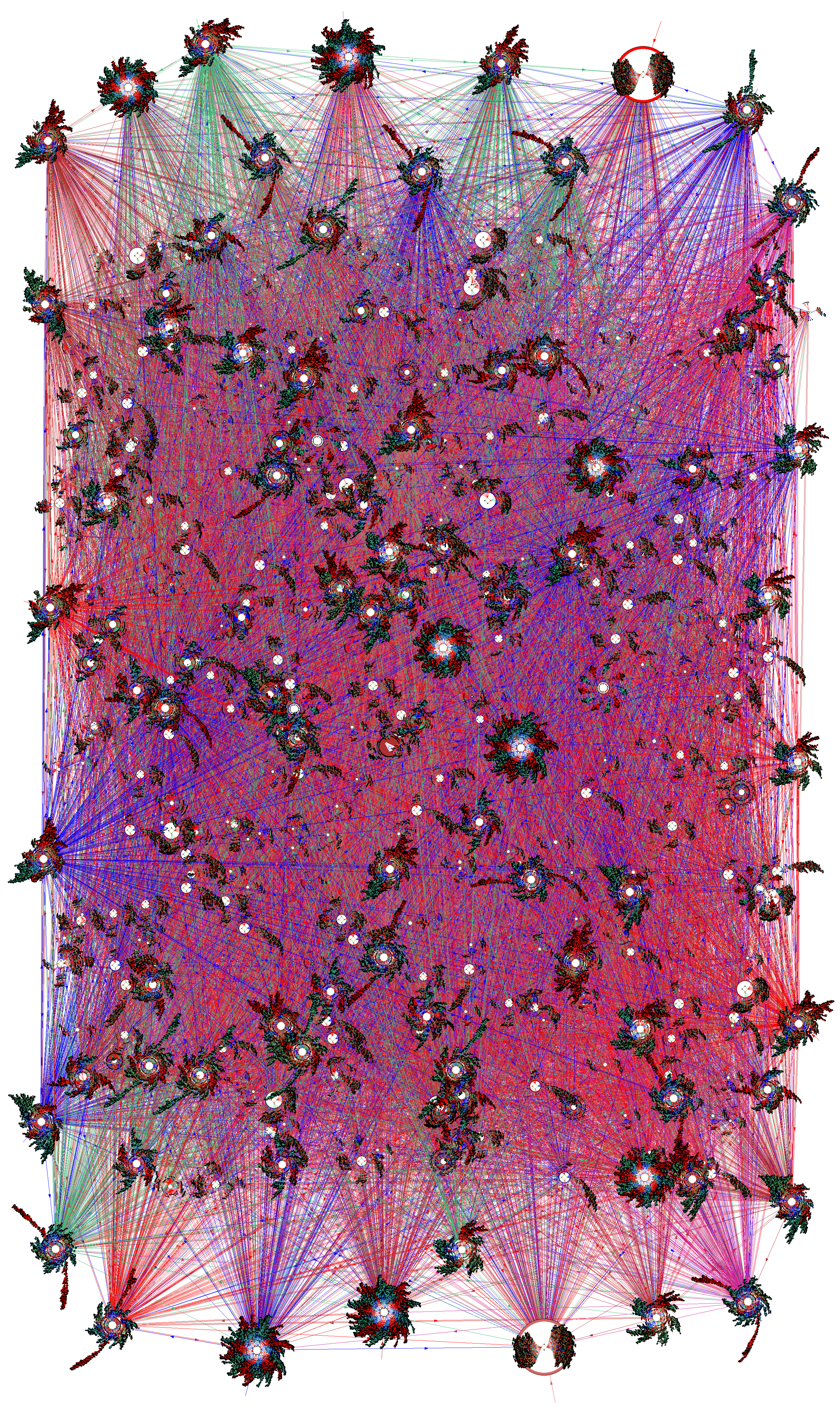}
\caption{Jump-graph composed from a basin field with configurations of 20 cells for ECA Rule 54. This meta-diagram consists of 428 basins yielding a field of 784,472 interconnections given by mutations of 1-bit value. Complex behaviour transits with a high density of connections between basins in the full graph for this automaton.}
\label{jumpGRule54ring20}
\end{figure}

\begin{figure}
\includegraphics[scale=.47]{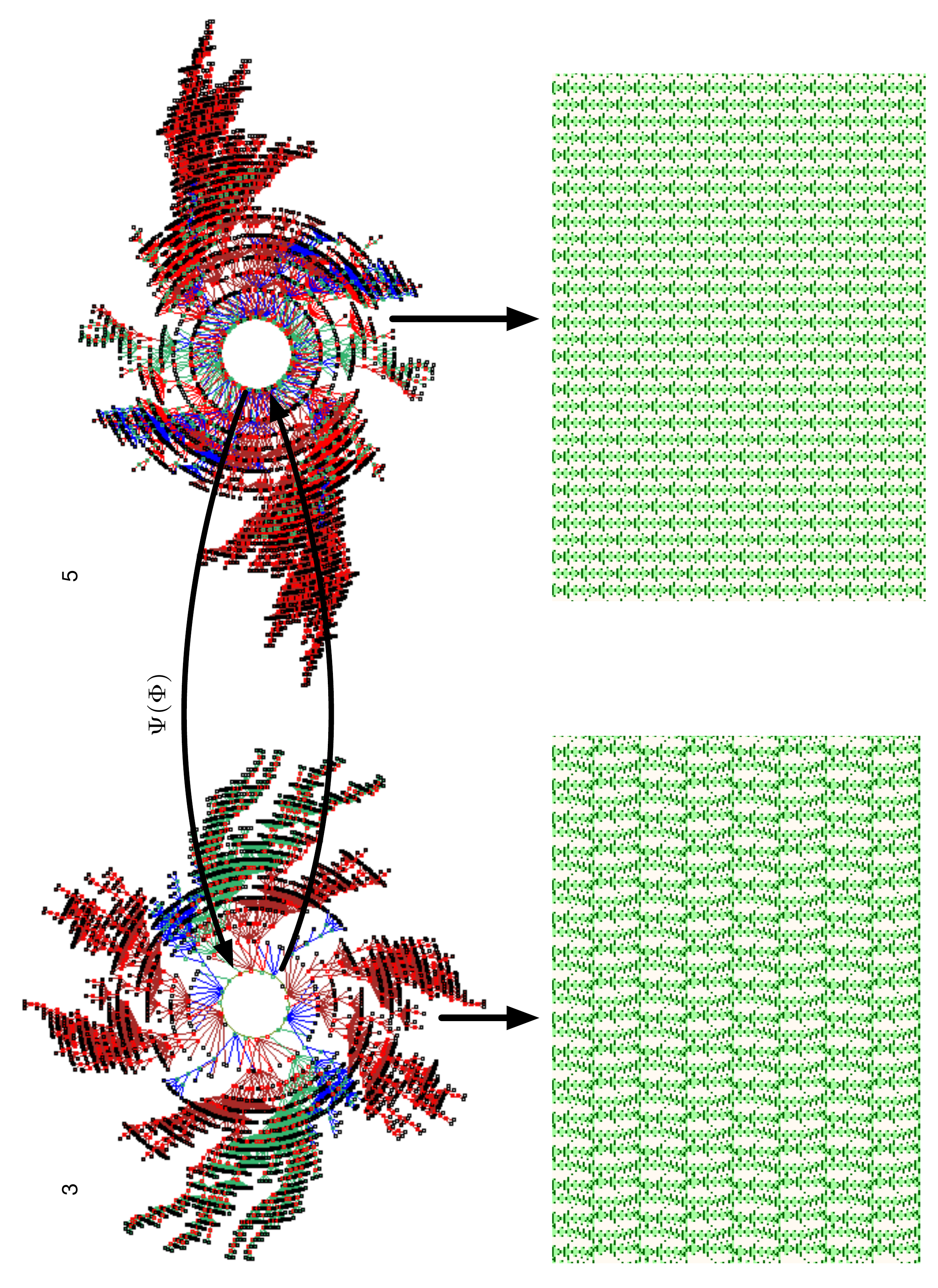}
\caption{A mutation between attractors characterised of a jump-graph in ECA Rule 54 (Fig.~\ref{jumpGRule54ring20}) for strings of length 20. The first basin (label with number three in DDLab) has a period 18 with a mass of 6,866 configurations. The second basin (label with number five) has a period 86 with a mass of 9,946 configurations. Both basins display dynamics with stationary localisations and synchronisations of small mobile localisations colliding in ECA Rule 54. Basins and evolutions are rotated 90 degrees. A filter is selected for a better view.}
\label{ECAR54mutationb3b5}
\end{figure}

In a jump-graph the question is: What is the probability that a string $w_i$ could mutate to another string $w_j$? Which could induce a change of transition to another attractor or itself (a loop, without change). This way, a configuration can remains in its attractor or jump to another attractor. We could find connections that jump to other attractor and later back to the original attractor with other mutation (a cycle between basins).

\begin{equation}
\Psi(\Phi(c_i)) \rightarrow \Phi(c_j)
\end{equation}

Let us consider a one-bit value mutation~\cite{ddlabbook}. We have a configuration $c_i$ expressed as a string $w_i = a_0 a_1 \ldots a_{n-1}$, such than it can jump into other configuration $c_j$ expressed as a string $w_j = b_0 b_1 \ldots b_{n-1}$. Hence $a_i$ can mutate to one $b_i$, where each configuration $c$ belongs at the same field of attractors $\Psi$. Also, if $a_i = b_i$, it represents a loop in the same basin.

Figure~\ref{jumpGRule54ring20} shows a jump-graph for Rule 54 with configurations of length 20. This field is composed of 428 basins on a domain of $k^{20}$. Each node has at least one mutation and therefore they all are connected (it is a graph strongly connected), determining 784,472 interconnections or mutations. For this automaton we have no unreachable states, because all states are connected, consequently a global stability is achieved in this system. This meta-diagram is a local universe containing basins with volume from 62 to 29,990 configurations and they have periods oscillating between 4 to 86 configurations. Non-trivial structures emerge during these developments with dozens of mobile localisations colliding in long transients; this implies a frequent change of basins into this jump-graph.

Finally, Fig.~\ref{ECAR54mutationb3b5} shows a mutation in detail, extracted from the full jump-graph calculated previously in Fig~\ref{jumpGRule54ring20}. There is a mutation between two attractors and they determine a change of dynamics. This diagram displays a connection between two basis (the basin three and the basin five as enumerated in DDLab\footnote{Discrete Dynamics Lab \url{http://www.ddlab.org}.} \cite{ddlabbook}). The mutation is expressed by the string $x0001011111101000100$ connecting both attractors where the change of dynamics implies a change of phase between stationary mobile localisations in Rule 54.

\section{Conclusions}

CA have very low entry fee --- anyone with basic coding skills can start experimenting with CA --- but very high exit fee --- far from anyone can produce results publishable in a reputable journal. This is because cell-start transition rules are appealingly simple yet space-time behaviour might be shockingly difficult to analyse and predict. There are few types of graphs/networks constructed over the cell-state transition rule spaces which might facilitate understanding of emergence and dynamics of complexity. They are de Bruijn diagrams, subsystems graphs, basins of attraction, and jump-graphs. De Bruijn diagrams characterise interrelationships between strings representation of the functions, they allows for some estimates of a distribution of states in a stable configuration. Subsystems graphs are determined by dynamics of travelling localisations, gliders, especially by interactions between mobile localisations. This is a promising representation especially when related to computational abilities, in a sense of collision-based computing, of CA rules supporting mobile localisations. Basins of attractions give us a straightforward visualisation of complexity: a number of cycles and heigh and bushiness of trees growing on them reflects sensitivity and `chaoticity' of the cell-state transition rules. Jump-graphs might be seen as characterising rules' tolerance to noise, mutations. There are two sides of the tolerance. Either a noise-tolerant system is very dull and therefore difficult to move to another loci of a global state space, or the system is so sophisticated that it have an in-build `mechanism' for dealing with mutation.

\end{document}